\documentclass[prb,showpacs,preprintnumbers,amsmath,amssymb,twocolumn]{revtex4-1}

\usepackage{graphicx}
\usepackage{dcolumn}
\usepackage{bm}
\usepackage{amsmath}
\usepackage{amssymb}

\usepackage{epstopdf}
\usepackage{gensymb}
\usepackage{array,multirow}
\usepackage{multirow}
\usepackage{dcolumn}
\usepackage{verbatim}
\usepackage{afterpage}
\usepackage{xfrac}
\usepackage{xcolor}
\usepackage[sort&compress]{natbib}
\usepackage[colorlinks=true,allcolors=blue]{hyperref}

\usepackage{latexsym}
\usepackage{graphicx}

\newcommand{\figwii}{0.65 in}
\newcommand{\figwiii}{1.55 in}
\usepackage{subfigure}
\usepackage{epsfig}
\usepackage{float}

\begin{document}


\title{Electronic Structure and Magnetism in the Layered Triangular Lattice Compound CeAuAl$_4$Ge$_2$}

\author{S. Zhang$^{1,2}$, N. Aryal$^{1,2}$, K. Huang$^1$, K.-W.Chen$^{1,2}$, Y. Lai$^{1,2}$, D. Graf$^1$, T. Besara$^1$, T. Siegrist$^{1,3}$, E. Manousakis$^{1,2,4}$, R. E. Baumbach$^{1,2}$
}
\affiliation{$^1$National High Magnetic Field Laboratory, Florida State University}
\affiliation{$^2$Department of Physics, Florida State University}
\affiliation{$^3$Department of Chemical and Biomedical Engineering, FAMU-FSU College of Engineering}
\affiliation{$^4$Department of Physics, National and Kapodistrian University of Athens, Athens, Greece}
\date{\today}

\begin{abstract}
Results are reported for the $f$-electron intermetallic CeAuAl$_4$Ge$_2$, where the atomic arrangement of the cerium ions creates the conditions for geometric frustration. Despite this, magnetic susceptibility measurements reveal that the low temperature magnetic exchange interaction is weak, resulting in marginally frustrated behavior and ordering near $T_{\rm{M}}$ $\approx$ 1.4 K. This occurs within a metallic Kondo lattice, where electrical resistivity and heat capacity measurements show that the Kondo-driven electronic correlations are negligible. Quantum oscillations are detected in ac-magnetic susceptibility measurements and uncover small charge carrier effective masses. Electronic structure calculations reveal that when the experimentally observed antiferromagnetic exchange interaction and the on-$f$-site Coulomb repulsion (Hubbard) U are considered, the $f$-electron bands move away from the Fermi level, resulting in electronic behavior that is dominated by the $s$-, $p$-, and $d$- bands, which are all characterized by light electron masses. Thus, CeAuAl$_4$Ge$_2$ provides a starting point for investigating geometric magnetic frustration in a cerium lattice without strong Kondo hybridization, where calculations provide useful guidance.
\end{abstract}

\pacs{Valid PACS appear here}
\maketitle


\section{\label{sec:level1}INTRODUCTION}
An attractive feature of correlated electron metals is that they often host multiple nearly degenerate electronic ground states, including charge order, spin order, multipolar order, superconductivity, heavy fermion behavior, breakdown of the Fermi liquid, and other intriguing phenomena\cite{Baumbach_JLTP_2010,Pfleiderer_RMP_2007,Hilbert_RMP_2007,Stewart_RMP_2001}. For materials containing $f$-electron elements with unstable valences (e.g. Ce and Yb), the behavior is associated with the competition between the Kondo hybridization and RKKY interactions\cite{Iglesias_PRB_1997,Doniach_PBC_1977}. However, rich behavior can also be driven by magnetic frustration which promotes complex ordering and might even lead to a quantum spin liquid state under some circumstances\cite{Si_PBCM_2006,Coleman_JLTP_2010,Vojta_PRB_2008,Balents_N_2010}. While this is an intriguing possibility, there is currently a lack of model systems to study; particularly for metals. Clearly then, it is desirable to uncover new $f$-electron metals with conditions for magnetic frustration.

Examples of $f$-electron materials with complex magnetism include YbRh$_2$Si$_2$\cite{Custers_PRL_2010}, CeRhIn$_5$\cite{Das_PRL_2014}, YbAgGe\cite{Sengupta_PRB_2010}, CePdAl\cite{Fritsch_PRB_2014}, CeRhSn\cite{Tokiwa_SA_2014}, and others. In tetragonal YbRh$_2$Si$_2$ and CeRhIn$_5$ the magnetic complexity presumably arises from fine balancing of the nearest neighbor, next nearest neighbor, etc. RKKY mediated interactions, which might be described in terms of a picture like the anisotropic next nearest neighbor Ising (ANNNI) model\cite{Selke_PR_1988}. In YbAgGe and CePdAl, magnetic frustration arises from the geometric arrangement of the $f$-electron ions in the crystalline lattice. Despite the differing origins for their magnetic behavior, these compounds are connected by the presence of strong hybridization between the $f$-electron and conduction electron states through the Kondo interaction. While this is often desirable, since strong hybridization can promote novel behavior, it also complicates the task of isolating which phenomena are primarily associated with magnetism. Thus it is important to find model systems where the hybridization strength is weak. Work on compounds such as $Ln$$T_2$$X_2$$M$ ($Ln$ = lanthanide, $T$ = Fe,Ru,Os, $X$ = Al, Ga, and $M$ = C,B)\cite{Baumbach_PRB_2012,Baumbach_JPCM_2012} where ANNNI-type magnetic frustration may shape the ordered ground state has made some progress in this direction, but it would be instructive to investigate other weakly correlated $f$-electron materials with the simpler condition of being geometrically frustrated.

Here we present results for single crystalline CeAuAl$_4$Ge$_2$, which forms in a layered rhombohedral structure with well separated planes of trivalent cerium ions that are distributed on a geometrically frustrated triangular lattice\cite{Ramirez_ARMS_1994}. The magnetic susceptibility is anisotropic, with an antiferromagnetic exchange interaction along the $c$-axis and a ferromagnetic in-plane exchange interaction that would not result in magnetic frustration. Nonetheless, crystal electric field splitting of the Hund's rule multiplet modifies the low temperature magnetism and results in small but negative Curie-Weiss temperatures for both directions, possibly setting the stage for weak geometric magnetic frustration. From specific heat and electrical resistivity measurements, we find magnetic order near $T_{\rm{M}}$ $\approx$ 1.4 K. Interestingly, we also observe magnetic fluctuations developing at much higher temperatures, starting near 15 K.  Thermodynamic, electrical transport, and quantum oscillation measurements reveal that the Kondo hybridization strength is weak, resulting in simple metallic behavior. Electronic structure calculations further reveal that by including the on-$f$-site Coulomb repulsion (Hubbard) U and the experimentally observed antiferromagnetic exchange interaction, the $f$-electron bands move away from the Fermi level. This results in electronic behavior that is dominated by the $s$-, $p$-, and $d$- bands which are all characterized by light electron masses. The calculated Fermi surface and associated quasi-particle effective masses are straightforward to understand in terms of quantum oscillations that are seen in ac-magnetic susceptibility measurements. From these results, we show that CeAuAl$_4$Ge$_2$ is a useful starting point from which to uncover phenomena resulting from geometric magnetic frustration in an weakly correlated $f$-electron metal, where calculations provide quantitative guidance.

\section{\label{sec:level1}EXPERIMENTAL DETAILS}
Single crystals of CeAuAl$_4$Ge$_2$ were grown using elements with purities $>$99.9\% in a molten Al flux. The starting elements were loaded into a 2 mL alumina crucible in the ratio 1(Ce):1(Au):10(Al):5(Ge). The crucible was sealed under vacuum in a quartz tube, heated up to 1000$^{\circ}$C at a rate of 83$^{\circ}$C/hr, kept at 1000$^{\circ}$C for 15 hours, and then cooled to 860$^{\circ}$C at a rate of 7$^{\circ}$C/hr, held at 860$^{\circ}$C for 48 hours, followed by cooling down to 700$^{\circ}$C at a rate of 12$^{\circ}$C/hr. After removing the excess flux by centrifuging the tubes, single-crystal platelets with typical dimensions of several millimeters in width and height and 0.5-1 mm thickness were collected. The crystals form as trigonal plates, where single crystal x-ray diffraction shows that the $c$-axis is perpendicular to the hexagonal plane.

Sample composition and structure characterization were performed by single crystal x-ray diffraction at room temperature using an Oxford-Diffraction Xcalibur2 CCD system with graphite monochromated Mo$K\alpha$ radiation. Data was collected using $\omega$ scans with 1$\degree$ frame widths to a resolution of 0.5 \textrm{\AA}, equivalent to $2\theta=90\degree$. The data collection, indexation, and absorption correction were performed using the Rigaku Oxford Diffraction CrysAlisPro software~\cite{CrysAlisPro}. Subsequent structure refinement were carried out using CRYSTALS~\cite{Crystals}, employing Superflip~\cite{Superflip}, with starting parameters from the literature~\cite{Wu_JSSC_2005,Pearson} to refine the structure. The data quality allowed for a full matrix refinement against $F^2$, with anisotropic thermal displacement parameters of all atoms in the structure. The occupancies of all atoms were relaxed but deviated negligibly from full occupancy. A crystallographic information file (CIF) has been deposited with ICSD (CSD No. 431206)~\cite{ICSD}.

Magnetization $M(H,T)$ measurements were performed on a single crystal at temperatures $T=$ 1.8 - 300 K under an applied magnetic field of $H=$ 0.5 T and for $0<H<7$ T at a temperature of $T=$ 1.8 K for $H$ applied both parallel($\parallel$) and perpendicular($\perp$) to the $c$-axis using a Quantum Design VSM Magnetic Property Measurement System. The specific heat $C(T)$ was measured for $T=$ 0.5 - 20 K and the electrical resistivity $\rho(T)$ was measured for $T=$ 0.6 - 300 K using a Quantum Design Physical Property Measurement System. AC susceptibility measurements were performed on a CeAuAl$_4$Ge$_2$ single crystal using a superconducting magnet at the National High Magnetic Field Laboratory under a sweeping field of $H=$ 0 - 18 T for several $T$ and also for several different angles $\theta$, where $\theta$ = 0 is defined as $H \parallel$ c.

Electronic structure calculations were performed using the WIEN2k package in the Full Potential Linearized Augmented Plane Wave (FPLAPW) and localized orbital basis framework \cite{Schwarz_CPC_2002} within the PBE parametrization of the generalized  gradient approximation (GGA) \cite{Perdew_PRL_1996}. The plane-wave cutoff parameter R$_{MT}$K$_{max}$ was chosen to be 7  where R$_{MT}$ is the muffin-tin radius and K$_{max}$ is the maximum size of the reciprocal lattice vectors.

\section{\label{sec:level1}RESULTS}

\begin{figure}[!t]
    \begin{center}
        \includegraphics[width=0.8\columnwidth]{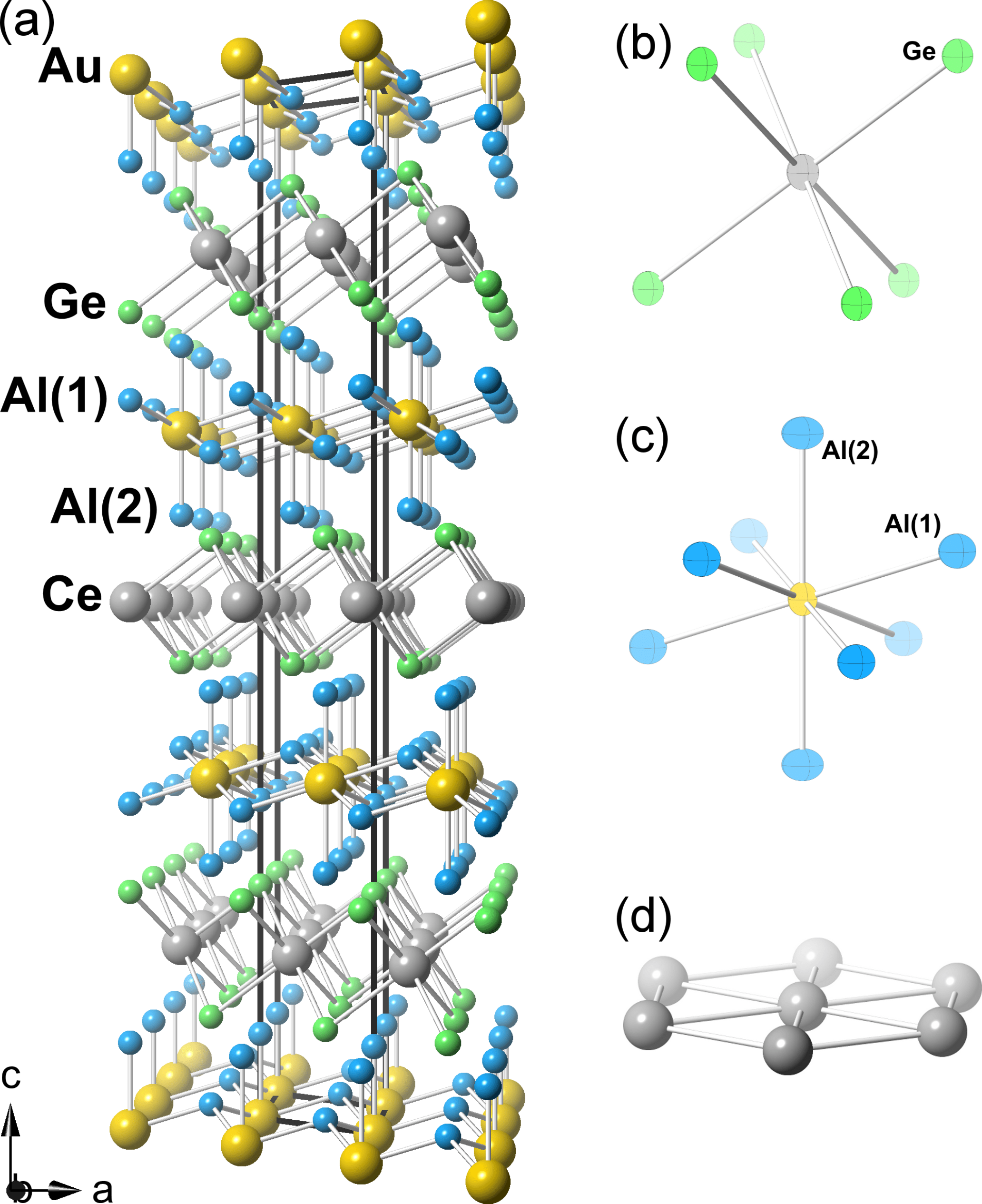}
        \caption{(a) Crystal structure of CeAuAl$_4$Ge$_2$. (b) and (c) Local environment of Ce and Au displaying anisotropic displacement ellipsoids with 95\% probability at 295 K. (d) Ce plane showing the equilateral triangle arrangement.}
        \label{fig:struct}
    \end{center}
\end{figure}

CeAuAl$_4$Ge$_2$ crystallizes in the rhombohedral space group $R\bar{3}m$ (No.166) with unit cell parameters $a=4.2334(2)$~\textrm{\AA} and $c=31.568(1)$~\textrm{\AA}. This structure (Fig.~\ref{fig:struct}(a)) -- explored in more detail by Wu and Kanatzidis~\cite{Wu_JSSC_2005} -- can be described as alternating layers of Ce and Au, stacked in an ...ABCABC... sequence. The Ce is octahedrally coordinated by Ge (Fig.~\ref{fig:struct}(b)), while Au is surrounded by Al forming a distorted cubic local environment (Fig.~\ref{fig:struct}(c)). Both the Ce-centered octahedra and the Au-centered cubes are edge-sharing. Focusing on the Ce layers, the Ce-atoms form a network of equilateral triangles in the $ab$-plane separated by the lattice constant $a$ (Fig.~\ref{fig:struct}(d)), and with the planes $1/3$ of the $c$-axis lattice constant apart ($d_{\rm{interplane}}$ $\approx$ 10.5 \AA), in essence creating a two-dimensional $f$-electron system where the triangular arrangement satisfies the condition for geometric frustration. Measurements for the crystals reported here show that the atomic site occupancy factors deviated negligibly from full occupancy, indicating a stoichiometric, high quality crystal. Further details of the XRD refinement and the structure are summarized in Table~\ref{tbl:xray}.

\begin{table}[b]
    \begin{center}
        \caption[]{Crystallographic data, and single crystal X-Ray diffraction collection and refinement parameters, obtained at ambient temperature. The bottom of the table lists atomic coordinates and equivalent thermal displacement parameters (in $\times10^{4}$ \textrm{\AA}$^2$).}
        \begin{tabular}{l c c c c c r}
            \hline
            \multicolumn{5}{l}{\textbf{Compound}} & \multicolumn{2}{l}{\textbf{CeAuAl$_4$Ge$_2$}} \\
            \hline
            \multicolumn{5}{l}{Formula weight (g/mol)} & \multicolumn{2}{l}{590.20} \\
            \multicolumn{5}{l}{Space group} & \multicolumn{2}{l}{$R\bar{3}m$ (No. 166)} \\
            \multicolumn{5}{l}{$a$ (\textrm{\AA})} & \multicolumn{2}{l}{4.2334(2)} \\
            \multicolumn{5}{l}{$c$ (\textrm{\AA})} & \multicolumn{2}{l}{31.568(1)} \\
            \multicolumn{5}{l}{$\rho_{\textrm{calc}}$ (g/cm$^{3}$)} & \multicolumn{2}{l}{6.000} \\
            \multicolumn{5}{l}{$Z$} & \multicolumn{2}{l}{3} \\
            \hline
            \multicolumn{5}{l}{} & \multicolumn{2}{l}{} \\
            \hline
            Atom & Site & Occ. & $x$ & $y$ & $z$ & $U_{\textrm{eq}}$ \\
            \hline
            Ce & 3b & 1 & 0 & 0 & 0.5 & 87(5) \\
            Au & 3a & 1 & 0 & 0 & 0 & 74(3) \\
            Al(1) & 6c & 1 & 0 & 0 & 0.30849(9) & 104(7) \\
            Al(2) & 6c & 1 & 0 & 0 & 0.08206(8) & 119(5) \\
            Ge & 6c & 1 & 0 & 0 & 0.22528(2) & 80(3) \\
            \hline
        \end{tabular}
        \label{tbl:xray}
    \end{center}
\end{table}

\begin{figure}[!t]
    \begin{center}
        \includegraphics[width=1.0\columnwidth]{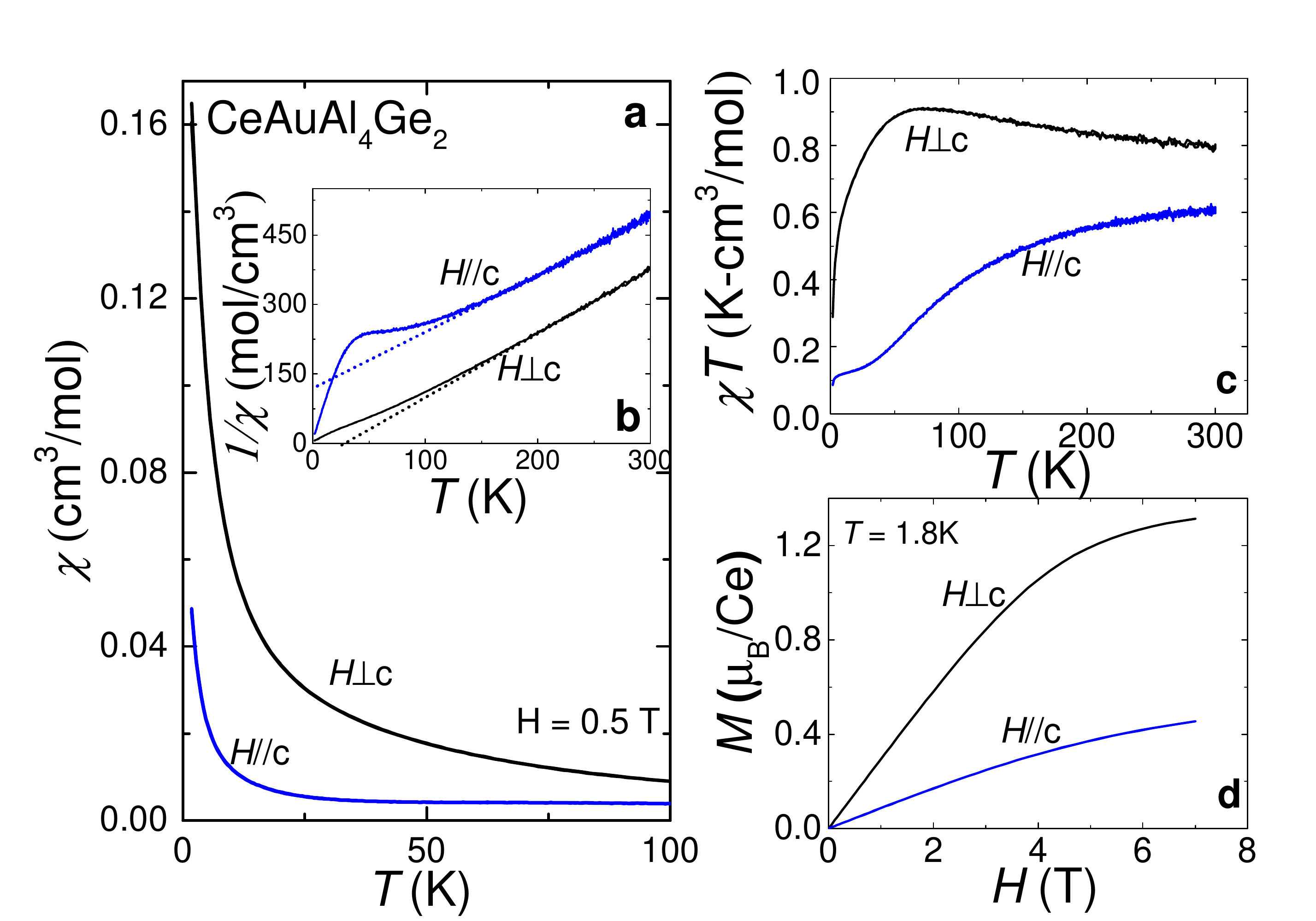}
        \caption{(a) Magnetic susceptibility $\chi(T)=M(T)/H$ for $H$=0.5 T
applied parallel and perpendicular to the $c$-axis vs temperature $T$ for CeAuAl$_4$Ge$_2$. (b) $\chi^{-1}(T)$ for $H \parallel$ c and $H \perp$ c. The dashed lines are Curie-Weiss fits to the data. (c) $\chi T(T)$ vs $T$ for $H \parallel$ c and $H \perp$ c. (d) Magnetization $M$ vs magnetic field $H$ for $H \parallel$ c and $H \perp$ c.}
        \label{fig:Magnetism}
    \end{center}
\end{figure}

\begin{figure}[!t]
    \begin{center}
        \includegraphics[width=1.0\columnwidth]{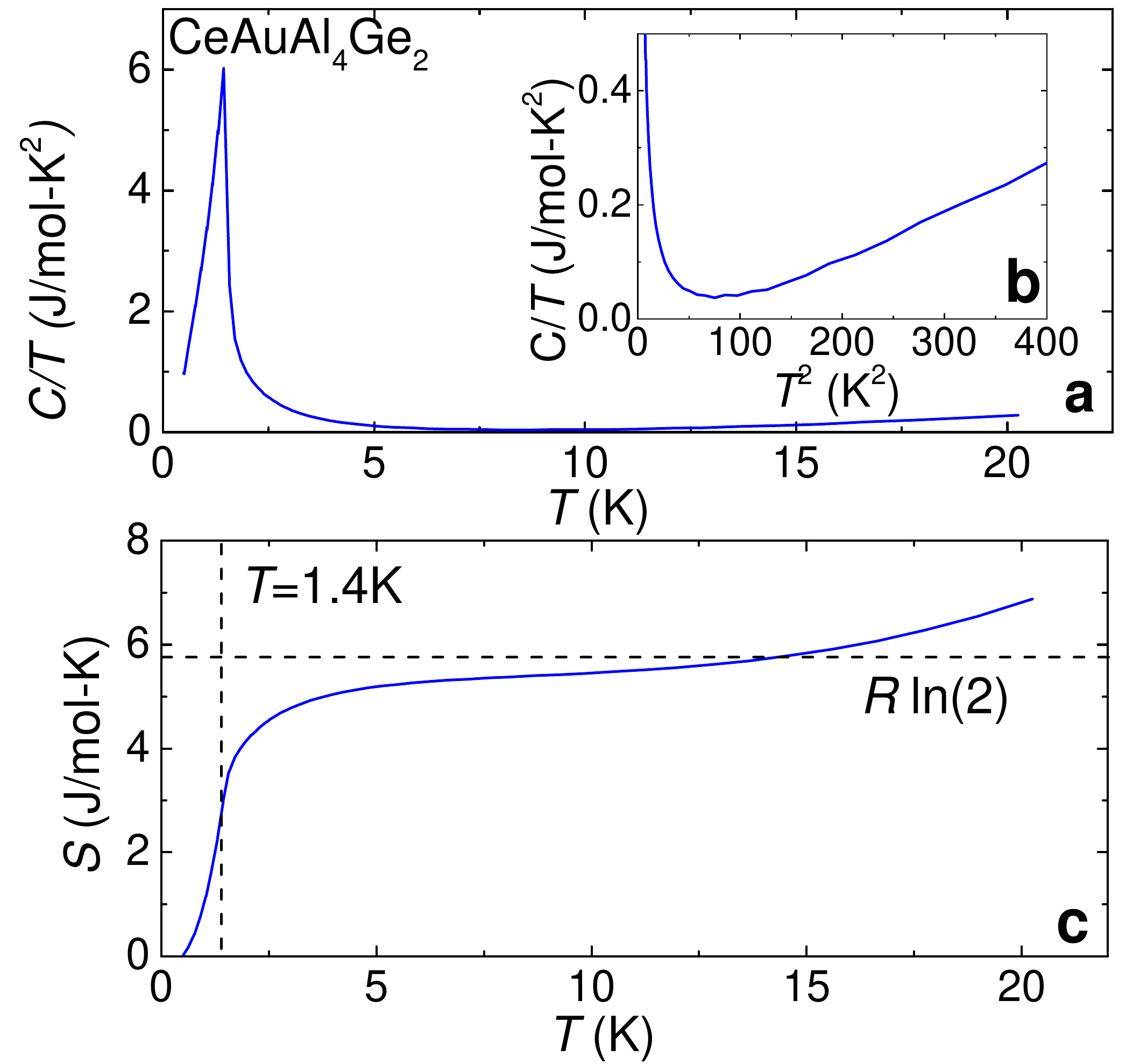}
        \caption{(a) Heat capacity divided by temperature $C/T$ vs $T$ for CeAuAl$_4$Ge$_2$. (b) $C/T$ vs $T^2$. (c) Entropy $S$ vs $T$. $S$ was calculated as described in the text.}
        \label{fig:C/T}
    \end{center}
\end{figure}

\begin{figure}[!t]
    \begin{center}
        \includegraphics[width=1.0\columnwidth]{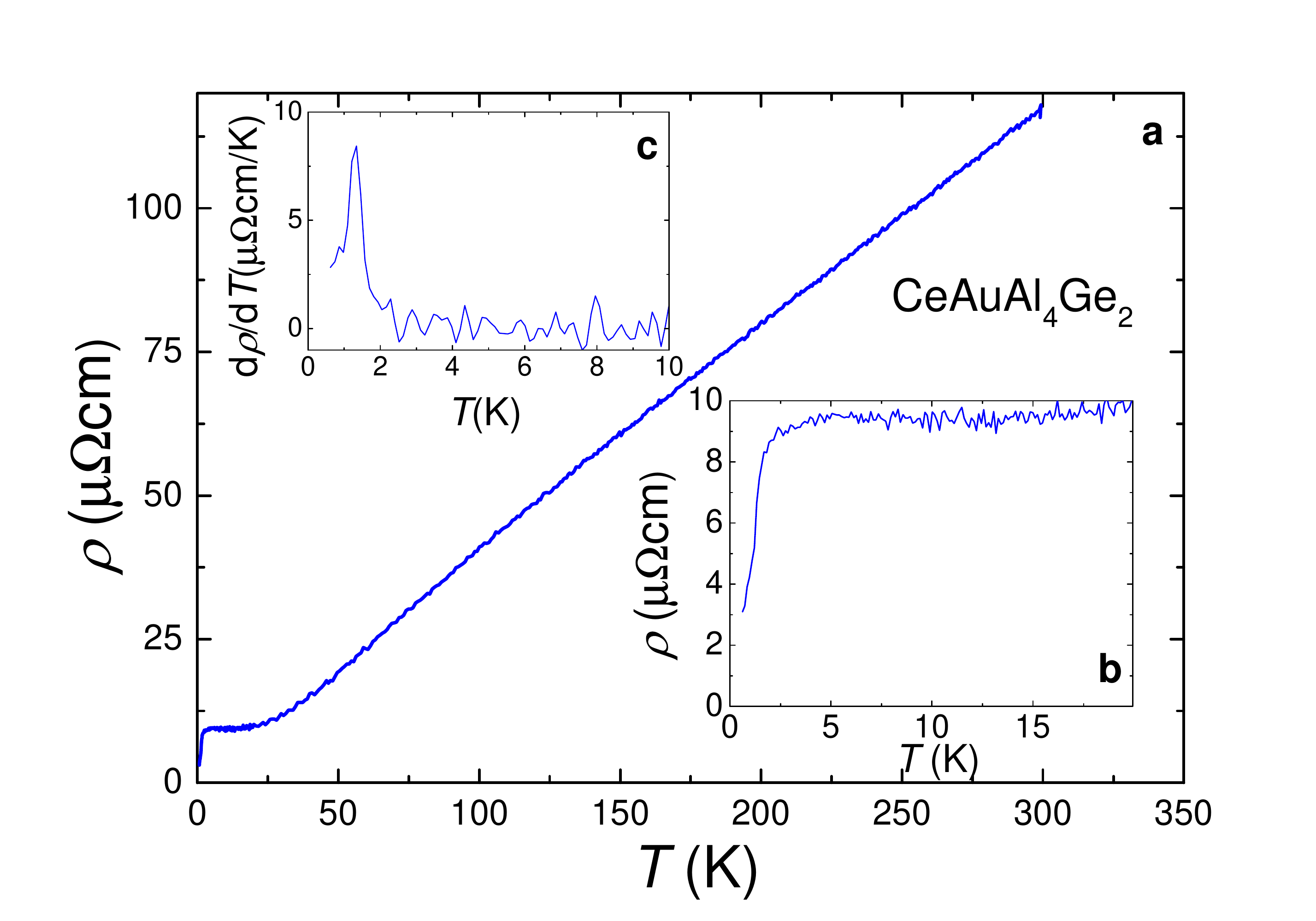}
        \caption{(a) Electrical resistivity $\rho$ vs temperature T for CeAuAl$_4$Ge$_2$. (b) Low temperature region of $\rho(T)$. (c) Derivative of the electrical resistivity with respect to temperature $\partial\rho/\partial T$ vs $T$ for $H$=0. The phase transition at $T_{\rm{M}}$ = 1.4 K appears as a sharp peak.
}
        \label{fig:rho}
    \end{center}
\end{figure}

The magnetic susceptibility data $\chi(T)=M(T)/H$ for CeAuAl$_4$Ge$_2$ in $H = 0.5$ T applied parallel ($\parallel$) and perpendicular ($\perp$) to the c axis are presented in Fig.~\ref{fig:Magnetism}. The data follow a Curie-Weiss (CW) temperature dependence for 150 K $\le T \le$ 300 K (Fig.~\ref{fig:Magnetism} (b)), given by the expression,\[\chi=\frac{C}{(T-\theta)} \eqno{(1)}\] where fits to the data yield $\theta$ = -90 K for $H\parallel c$ (24 K for $H\perp c$) and $C= 0.73$ ($0.79$) giving effective magnetic moments $\mu_{eff}\approx2.42 \mu_B$/Ce (2.51 $\mu_B$/Ce), close to the expected Hund's rule value ($\mu_{eff}= 2.54$ $\mu_B$/Ce). The anisotropy between the $H\parallel c$ and $\perp c$ curves reveals that the $ab$-plane is the easy axis for the magnetization, and that ferromagnetic and antiferromagnetic exchange interactions occur in $H \parallel$ and $H \perp c$ directions, respectively. These trends are emphasized in Fig.~\ref{fig:Magnetism}c where (1) $\chi T$ approaches 0.8 K-cm$^2$/mol at 300 K, as expected for trivalent Ce, and (2) the $\chi$$T$ curves for $H \parallel$ and $H \perp$ $c$-axis increase and decrease with decreasing $T$, consistent with ferromagnetic and antiferromagnetic exchange interactions, respectively. Thus, the high temperature Curie-Weiss behavior indicates that although the Ce atoms are arrayed such that their easy axis for magnetization lies in the plane of a geometrically frustrated lattice, they experience a ferromagnetic exchange interaction that would circumvent frustration.

With decreasing temperature, a shoulder appears in $\chi (T)$ near 25 K for $H\parallel c$ and $\perp c$ which is likely related to crystal electric field splitting of the cerium Hund's rule multiplet. Similar to what is seen in many other Ce-based magnets, this reduces the ground state magnetic moment and modifies the magnetic exchange interaction\cite{Christianson_PRB_2004}. This is evident in $M(H)$ curves at $T$ = 1.8 K for $H\parallel(\perp)$ $c$ (Fig.~\ref{fig:Magnetism}(d)), which behave as Brillouin functions and have saturation moments near $M_{sat}\approx 0.4 \mu_B$/Ce for $H\parallel c$ and $1.3 \mu_B$/Ce for $H\perp c$ above $H\approx$ 5 T, which are reduced from the full saturation moment of the J = 5/2 cerium multiplet. For $T < 20$ K, Curie-Weiss fits to the data yields $\theta$ = -0.9 K for $H \parallel c$ ( -2.8 K for $H \perp c$) and $C$ = 0.13 cm$^3$K/mol for $H \parallel c$ (0.78 cm$^3$K/mol for $H \perp c$). Here the value for $\theta$ in the $ab$-plane indicates an antiferromagnetic interaction, which sets the stage for magnetic frustration, albeit weak, at low temperatures. We further note that the small values of $\theta$ indicate that the low temperature Kondo temperature ($T_{\rm{K}}$) is small: according to Ref.\cite{Krishna-murthy_PRL_1975}, $T_{\rm{K}}$ $\approx$ $\theta$/2.

The specific heat divided by temperature $C/T$ data are shown in Fig.~\ref{fig:C/T}(a), where there is a large $\lambda$-like feature associated with the second order phase transition near $T_{\rm{M}}$ = 1.4 K, which is on the order of the low temperature Curie-Weiss constants from $\chi(T)$. In Fig.~\ref{fig:C/T}(b) we show $C/T$ vs. $T^2$ which, for a Fermi liquid, is expected to follow the expression $C/T$ = $\gamma$ + $\beta T^2$ for temperatures well below the Debye temperature $\theta$$_{\rm{D}}$. We find that there is no such region in this data set, possibly because the magnetic fluctuations that precede the ordered state extend up to $T \approx$ 10 K. This may be a signature of geometric frustration which spreads the magnetic entropy above $T_{\rm{M}}$. Although it is not possible to extrapolate $C/T$ vs $T^2$ to zero temperature to extract a value for $\gamma$, we estimate that the value is small ($\gamma \lesssim $ 15 mJ/mol-K$^2$). This further suggests that the hybridization between the $f$- and conduction electrons from the Kondo effect is weak. In Fig.~\ref{fig:C/T}(c) we show the entropy $S$ vs. $T$ obtained by integrating $C/T$ from 0.5 K. This quantity is only a rough estimate of the 4$f$ entropy (recall the data only reaches down to 0.49 K and the electronic contribution could not be isolated from the phonon contribution), but still provides some useful insight. While $S$ increases abruptly at $T_{\rm{M}}$ =1.4 K, the entropy only reaches 52\% of $R\ln2$ and recovers 90 \% of this value near $T=5$ K. This provides further evidence that magnetic fluctuations persist at temperatures above the ordering temperature, as might be expected if magnetic frustration influences the low temperature behavior.

The electrical resistivity $\rho$ vs $T$ data, where the current was applied in the ab plane, are shown in Fig.~\ref{fig:rho}. Starting from 300 K, $\rho(T)$ decreases monotonically as is expected for a typical metal, where the lattice contribution is the main term. Over the entire temperature range, there is no evidence for Kondo lattice behavior, indicating that CeAuAl$_4$Ge$_2$ is dissimilar from prototypical heavy fermion materials (e.g., CeRhIn$_5$\cite{Knebel_JPSJ_2008}, CeCu$_2$Si$_2$\cite{Yuan_S_2003}, and CeCu$_6$\cite{Stewart_PRB_1984}). Fig.~\ref{fig:rho}(b) highlights $\rho(T)$ near the magnetically ordered state, where the curve saturates to a constant value below $T \approx$ 15 K, after which it abruptly decreases below $T_{\rm{M}}$ = 1.4 K due to the removal of spin fluctuation scattering of conduction electrons. At low temperatures, $\rho(T)$ saturates towards a value $\rho_0$ $\approx$ 1.2 $\mu\Omega$cm, giving a residual resistivity ratio $RRR = \rho_{300k}/\rho_0 \approx 104$. This large value reveals the high quality of these specimens, and is consistent with results from x-ray diffraction. We note that the $RRR$ was determined for several different crystals from the same batch, with $RRR$ reaching values as high as 160.

\begin{figure}[!t]
    \begin{center}
        \includegraphics[width=1.0\columnwidth]{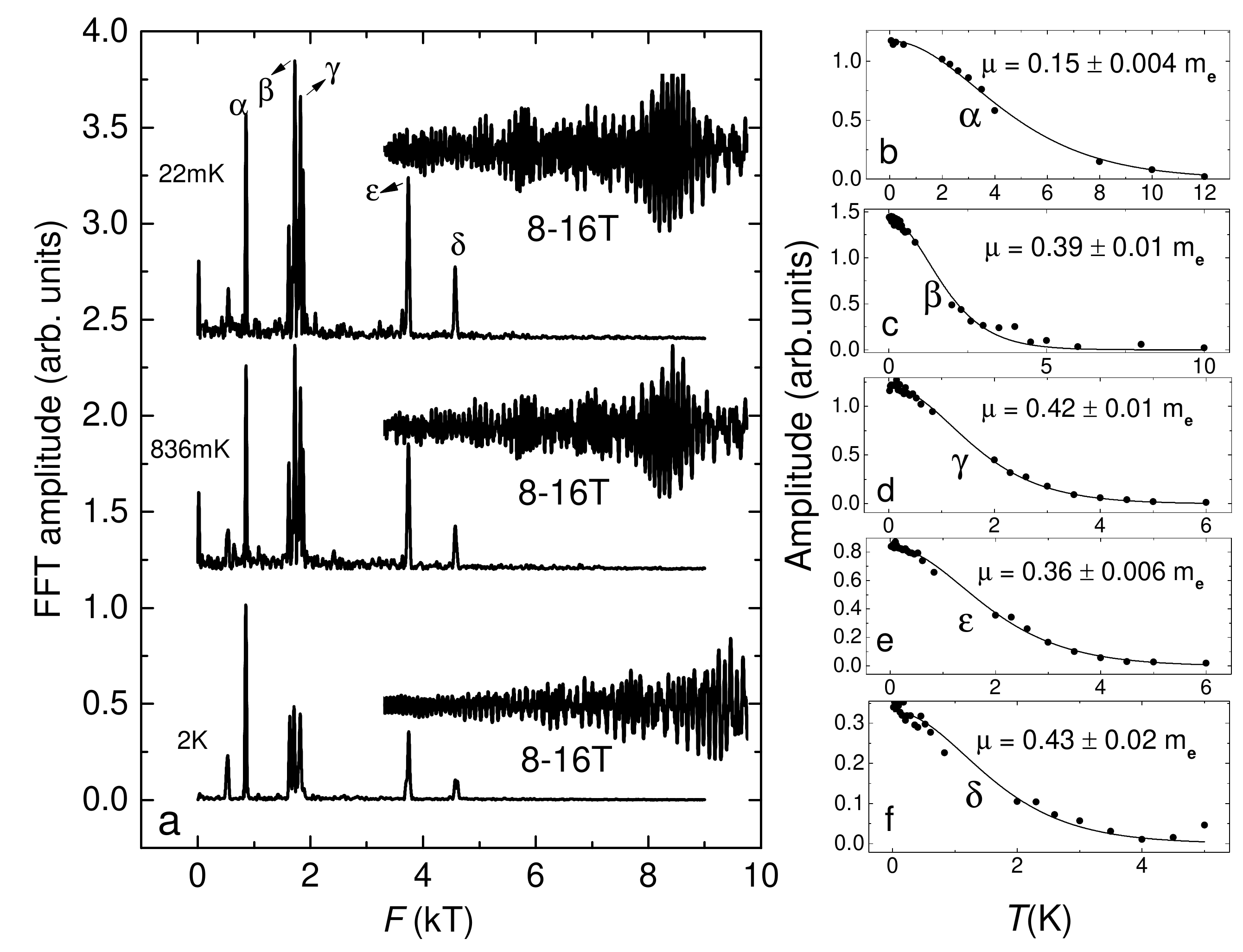}
        \caption{(a) Quantum oscillations observed in the ac-magnetic susceptibility vs magnetic filed $H$ and their Fast Fourier Transformations plotted vs frequency. The background was subtracted as described in the text. (b)-(f) Amplitude vs. temperature data fitted with the Lifshitz-Kosvech function.}
        \label{fig:QO}
    \end{center}
\end{figure}


In Fig.~\ref{fig:QO} we show results from ac-magnetic susceptibility measurements in magnetic fields up to $H$ = 16 T with $H \parallel c$ and for 20 mK $< T <$ 2 K. The monotonic background has been removed by subtracting a polynomial curve in order to reveal quantum oscillations, which first appear near 6 T. As evidenced by the small magnetic field required to observe the onset of quantum oscillations, the samples are of very high quality with an extremely low amount of defect scattering. A Fast Fourier transform (FFT) was performed in order to extract the different frequencies $F$ (Fig.~\ref{fig:QO} (a)). The Onsager relation $F = A(\Phi_0/2 \pi^2)$, where $\Phi_0$ = $h/2 e$ is the magnetic flux quantum, $h$ is the Planck constant and $e$ is the electron charge, relates the measured frequencies to the extremal  cross-sectional areas $A$ of the Fermi surface\cite{CUP}. For $H$ applied along the $c$-axis, there are several frequencies that are summarized in Table~\ref{table2}. The same frequencies are extracted regardless of the interval of $H^{-1}$ used for the FFTs. We focus on the temperature dependence of these frequencies as summarized in Fig.~\ref{fig:QO} (b)-(f). These data are well described by the expression\cite{Terashima_PRB_1997} for a Fermi liquid,
\[A \propto \sqrt{H} R_T R_D\]
 where $A$ is the amplitude of the oscillations, $R_D$ is the damping factor related to the Dingle temperature and the thermal damping factor $R_T$ $=$ $\frac{\alpha T m^* / H}{\sinh(\alpha T m^* / H)}$. In this expression, $m^*$ is the charge carrier effective mass and $\alpha$ $=$ 2$\pi^2$$k_B$$m_e$/$e$$\hbar$ $\approx$ 14.69 T/K where $m_e$ is free electron mass. Fits to the data yield small effective masses (Table~\ref{table2}), which are comparable to those typically seen in non-4$f$ containing analogues to prototypical heavy fermion materials: e.g., LaCoIn$_5$\cite{Hall_PRB_2009} and ThRu$_2$Si$_2$\cite{Matsumoto_JPSCP_2014}, again indicating that the 4$f$-electrons in CeAuGe$_4$Al$_2$ do not significantly hybridize with the conduction electrons.

\section{\label{sec:level1}DFT CALCULATIONS}

Electronic structure calculations were performed in order to clarify the magnetic ordering and quantum oscillation results. We first investigated the simplest case of ferromagnetic ordering (not shown). Regardless of whether spin orbit coupling is included, these calculations produce several flat bands near the Fermi level which have a significant $f$-electron character. Such bands result in enhanced effective mass charge carrier quasiparticles, which contradicts the experimental result. Motivated by this disagreement and since our experiments suggest antiferromagnetic ordering, we allowed for antiferromagnetic alignment by carrying out a DFT calculation using the conventional hexagonal cell instead of the primitive rhombohedral cell. An in-plane antiferromagnetic ordering requires having more than one Ce atoms in the ab-plane. Hence, the hexagonal cell along the x-direction was doubled and the spins were relaxed. In this case an added on Ce-$f$ Coulomb repulsion (Hubbard) U term has a significant effect on the massive Ce-$f$ character bands near the Fermi level:  It moves them away from the Fermi level and this explains the absence of QO frequencies with very massive electrons. The features and character of the other bands are largely unaffected by the inclusion of the Hubbard U term. This unit cell-structure has 48 atoms in the unit cell and makes our calculation computationally demanding. For the self-consistent calculation, we used a $k-$point mesh of $5\times 11\times 4$.

\begin{figure}[htp]
    \begin{center}
        \subfigure[]{
            \includegraphics[width=\figwiii]{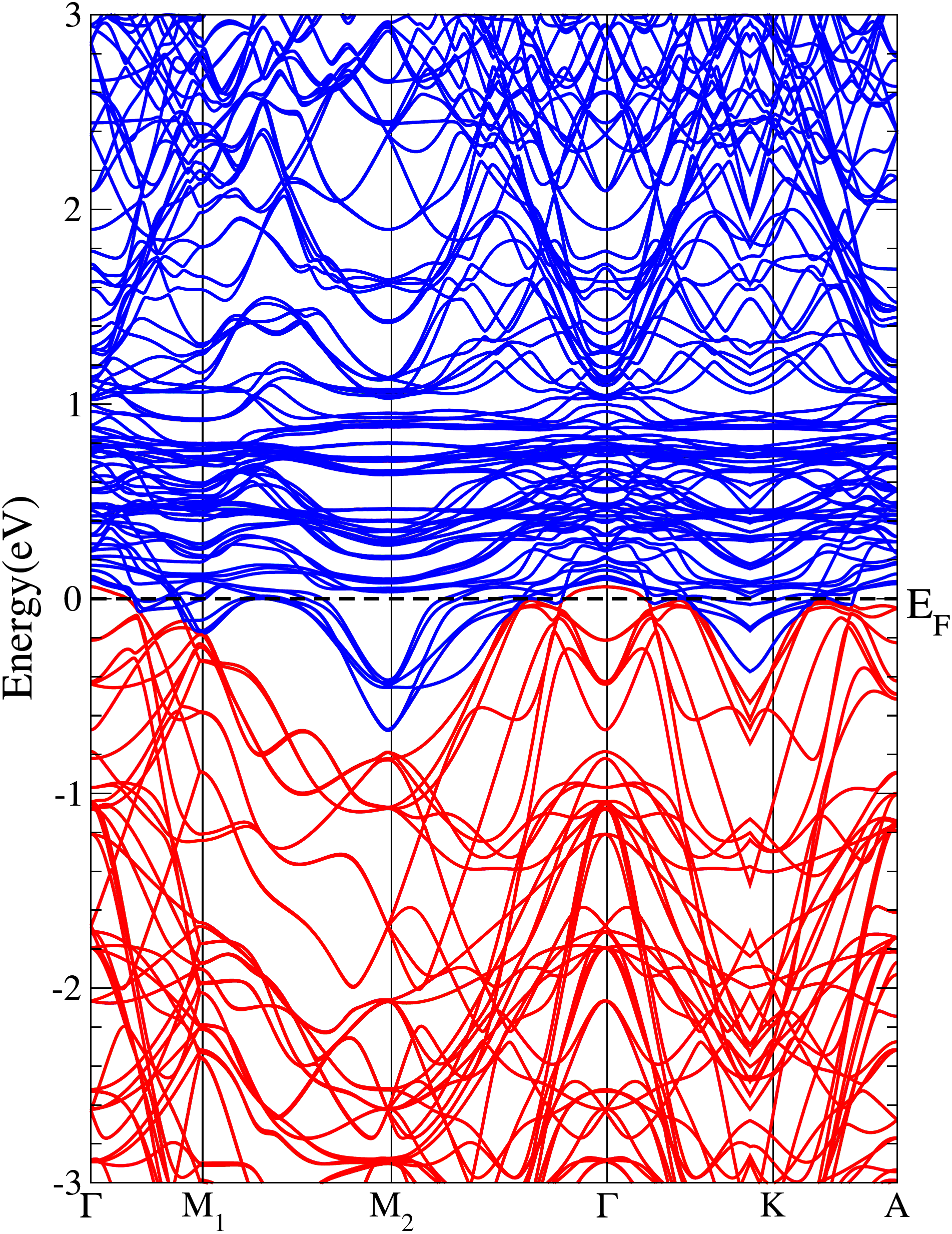}
            \label{fig:CeAlAuGe_nsoc_noU}
        }
\hskip 0.0 in
        \subfigure[]{
            \includegraphics[width=\figwiii]{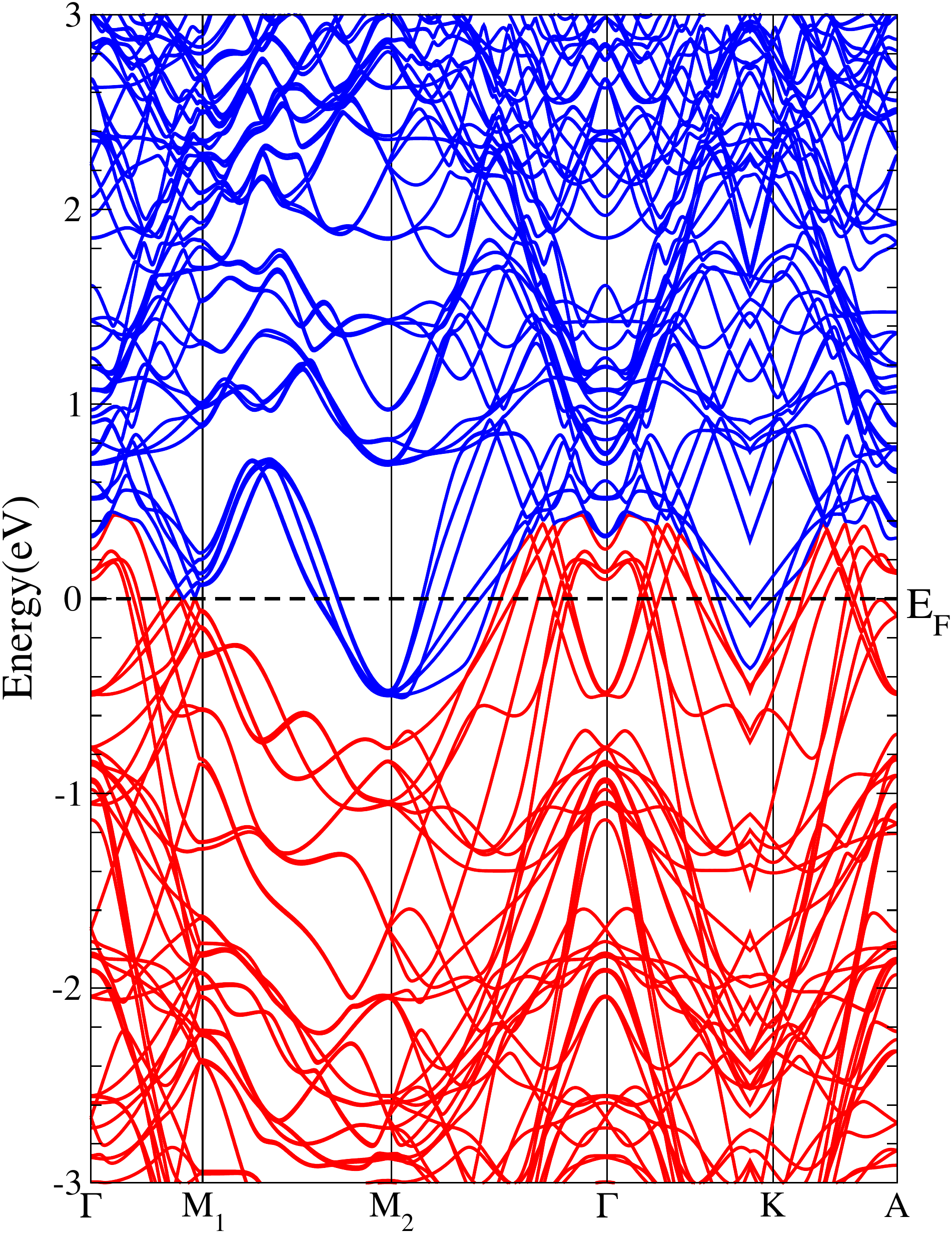}
            \label{fig:CeAlAuGe_soc}
        }\\


    \end{center}
        \caption{
          Band Structure in the antiferromagnetic phase of CeAuAl$_4$Ge$_2$ along the direction $\Gamma$-M$_1$-M$_2$-$\Gamma$-K-A where M$_1$ = (1/2,0,0), M$_2$ = (0,1/2,0), K = (2/3,-1/3,0), and A = (0,0,1/2) of the rectangular BZ. (a) Without U and without SOC and (b) With U=6.80 eV and J=0.680 eV and including SOC.
}
    \label{fig:bands_AF}

\end{figure}

\begin{figure}[!t]
    \begin{center}
        \includegraphics[width=1.0\columnwidth]{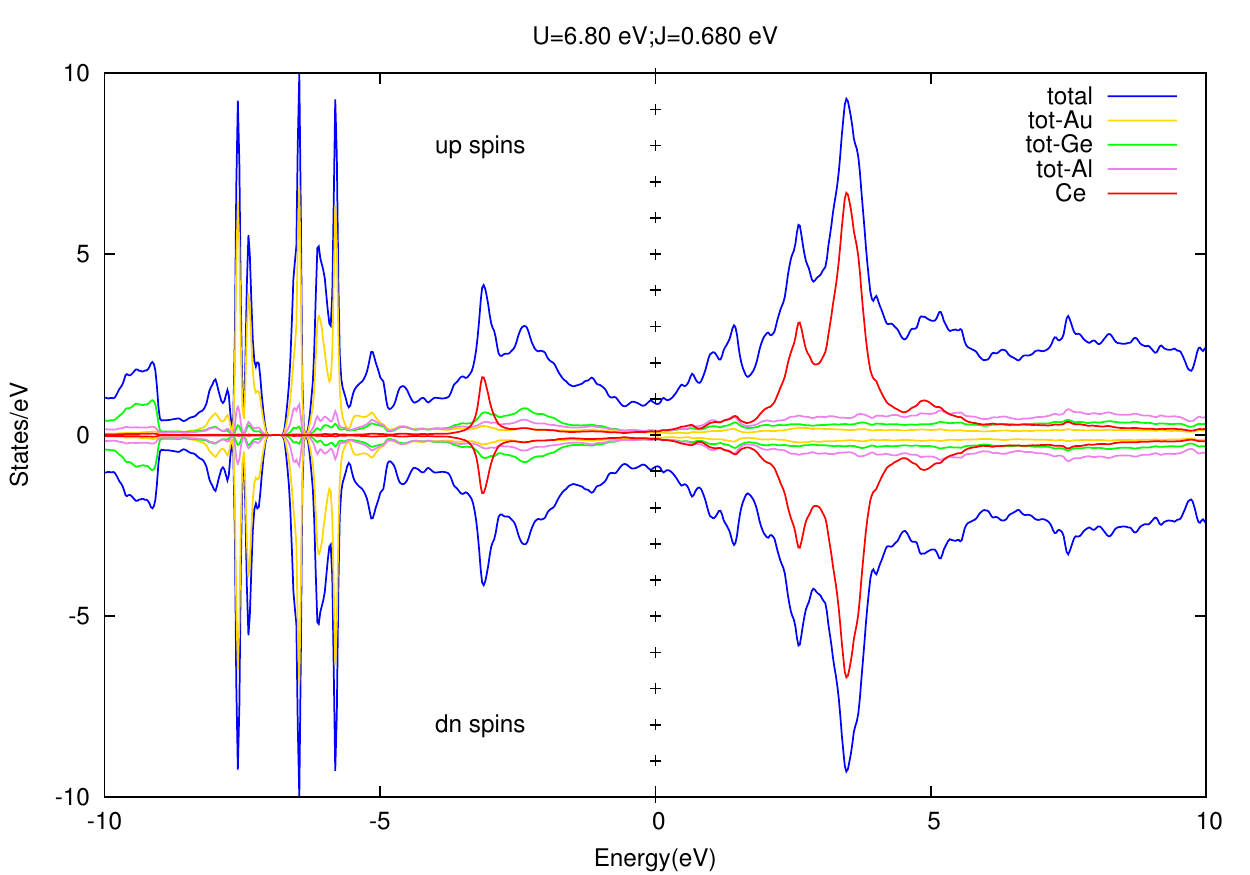}
        \caption{Projected density of states for CeAuAl$_4$Ge$_2$ where the antiferromagnetic exchange and on $f$-site Coulomb repulsion (Hubbard) U are considered, as described in the text.}
        \label{fig:big}
    \end{center}
\end{figure}

\begin{figure}[htp]
    \begin{center}
        \subfigure[]{
            \includegraphics[width=\figwii]{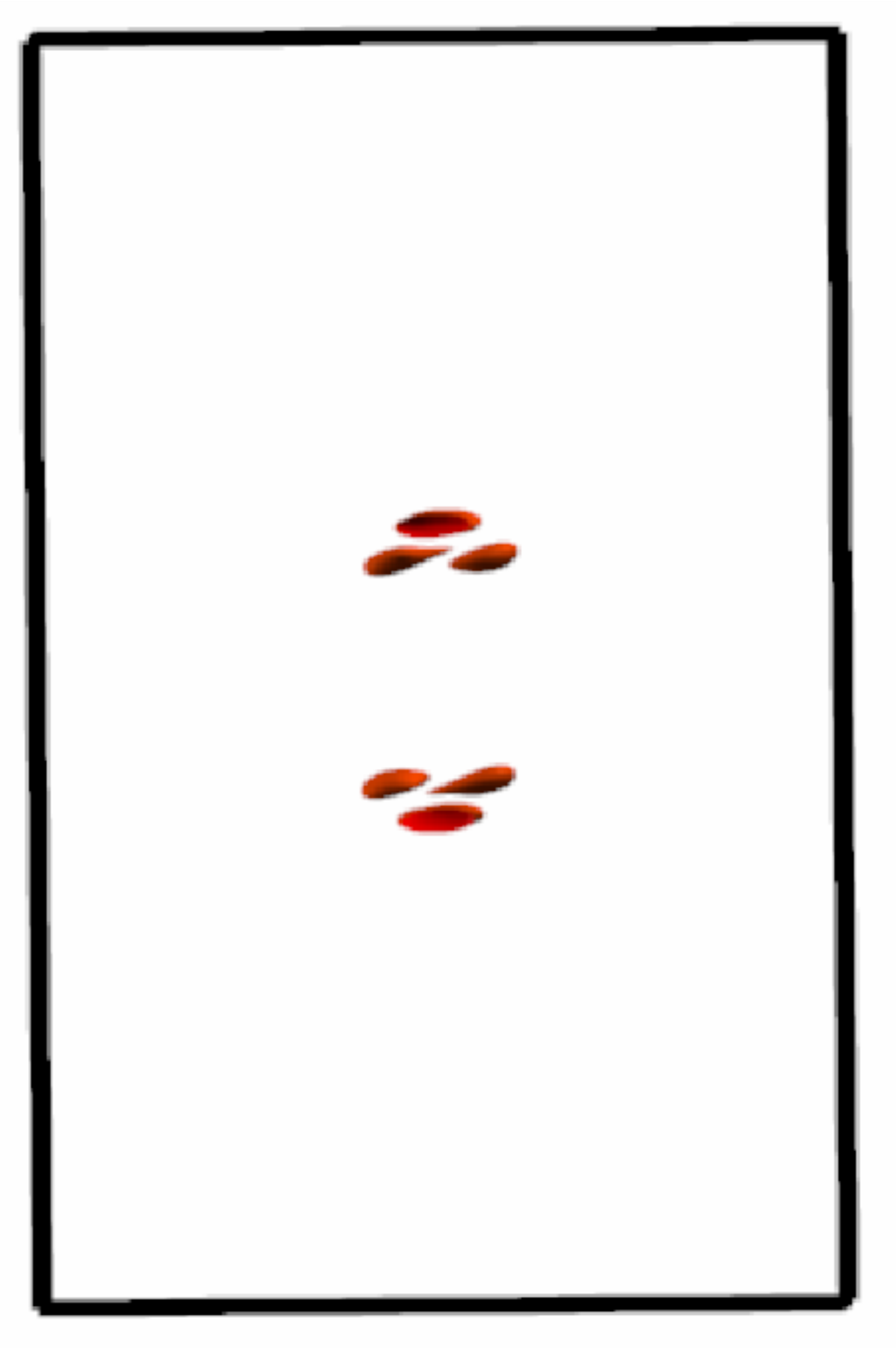}
            \label{fig:band_1}
        }
\hskip 0.1 in
        \subfigure[]{
            \includegraphics[width=\figwii]{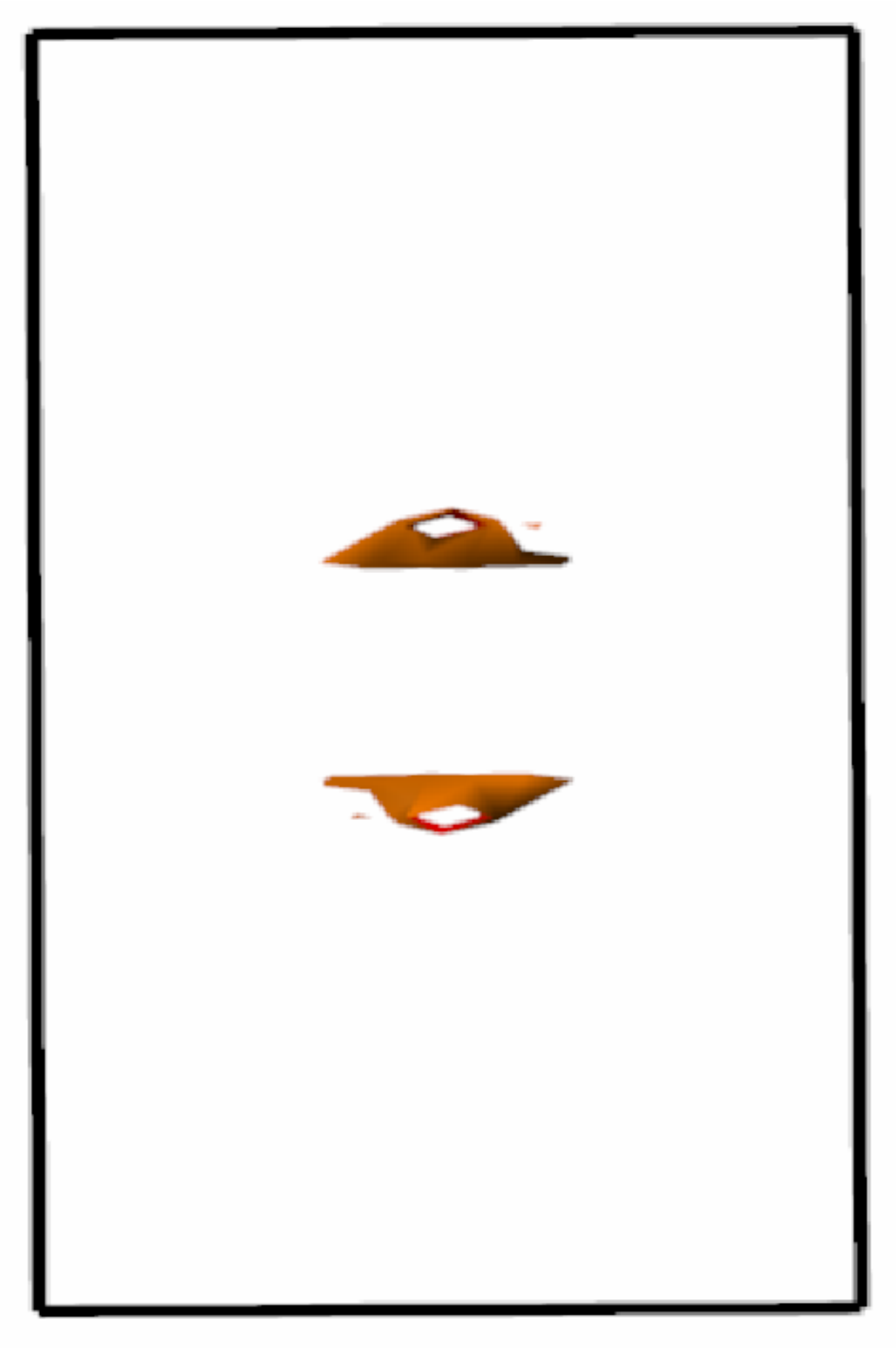}
            \label{fig:band_2}
        }
\hskip 0.1 in
        \subfigure[]{
            \includegraphics[width=\figwii]{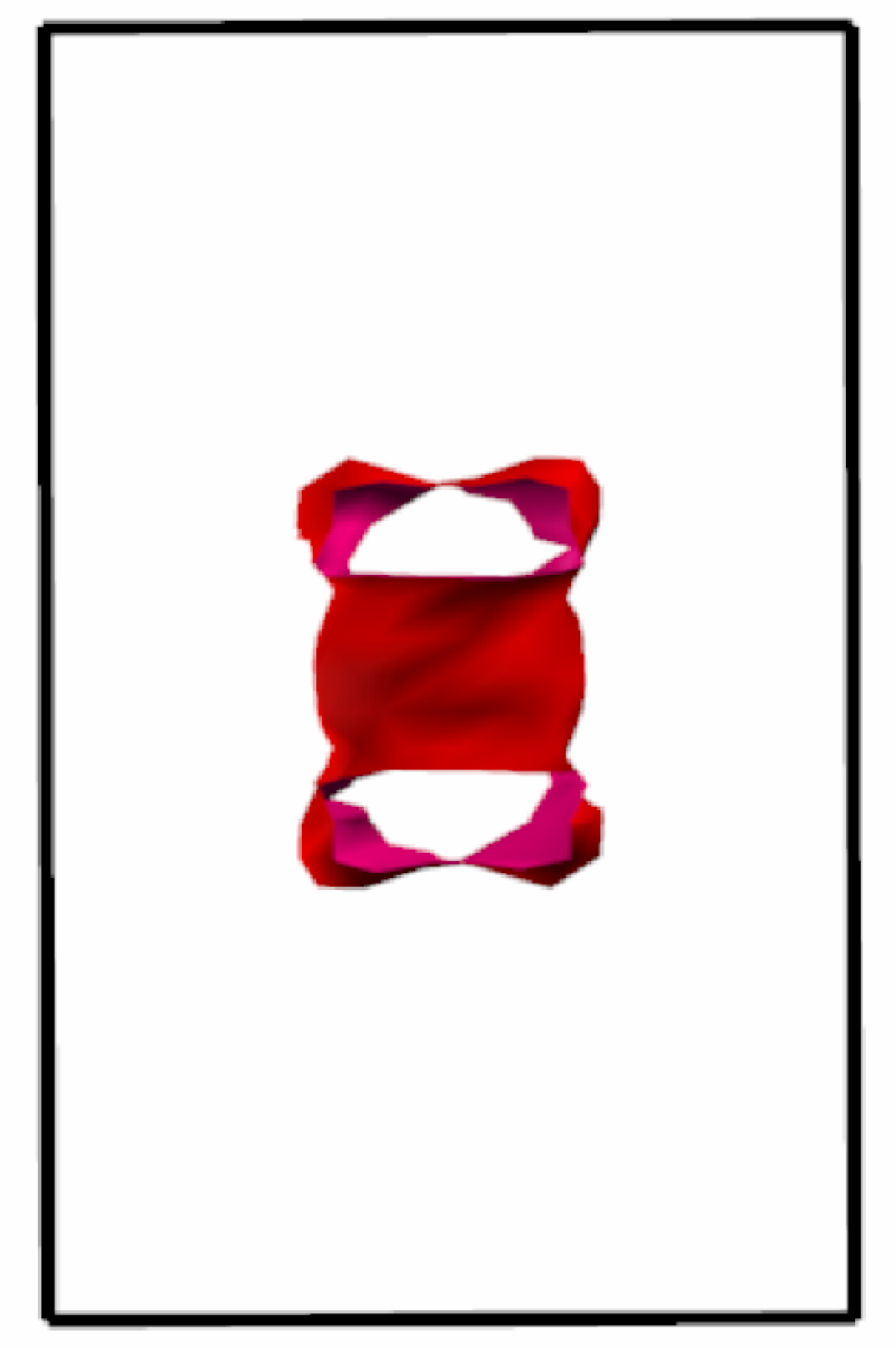}
            \label{fig:band_3}

        }
\hskip 0.1 in
        \subfigure[]{
            \includegraphics[width=\figwii]{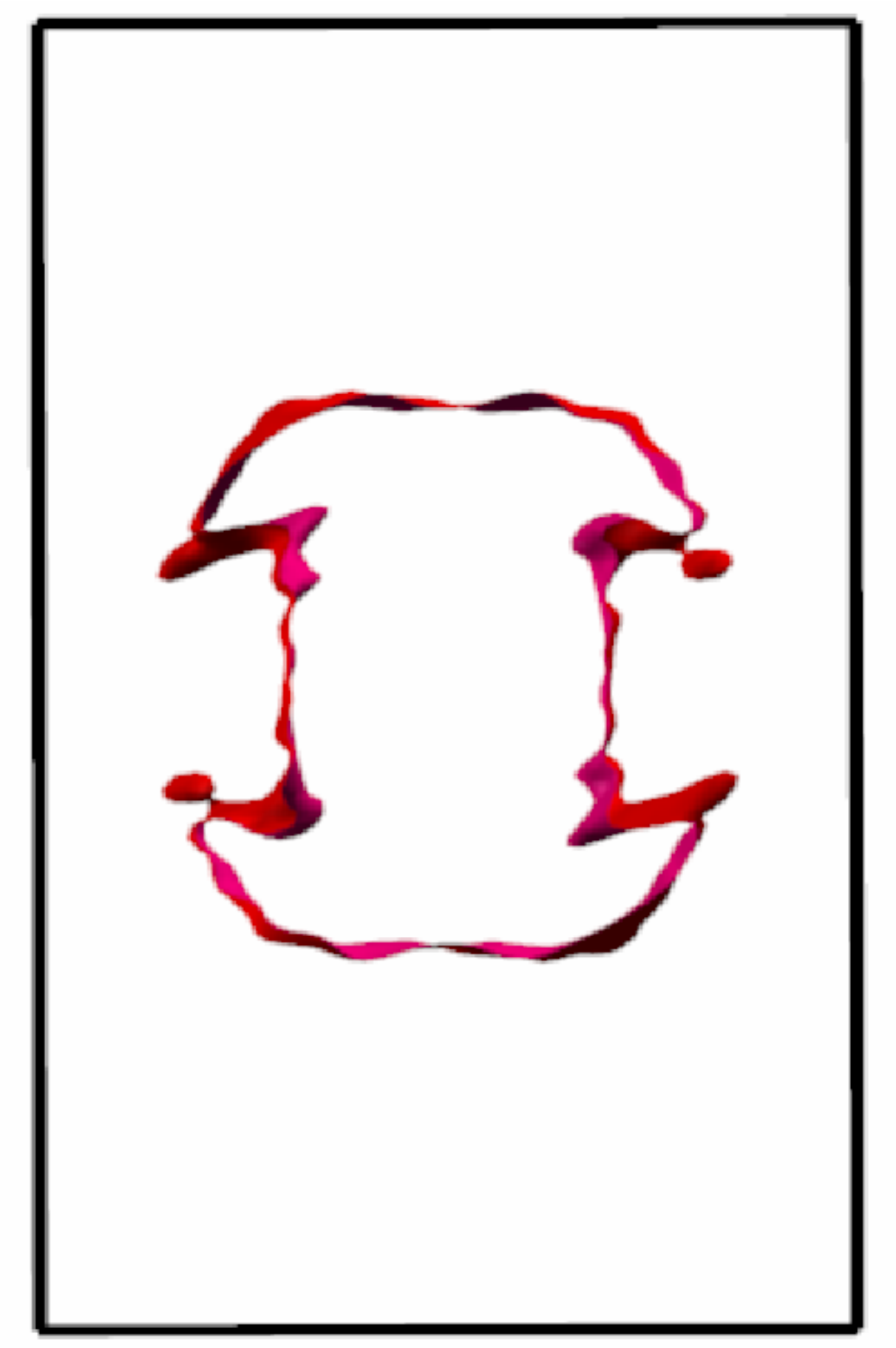}
            \label{fig:band_4}
        }\\

\vskip 0.1 in
        \subfigure[]{
            \includegraphics[width=\figwii]{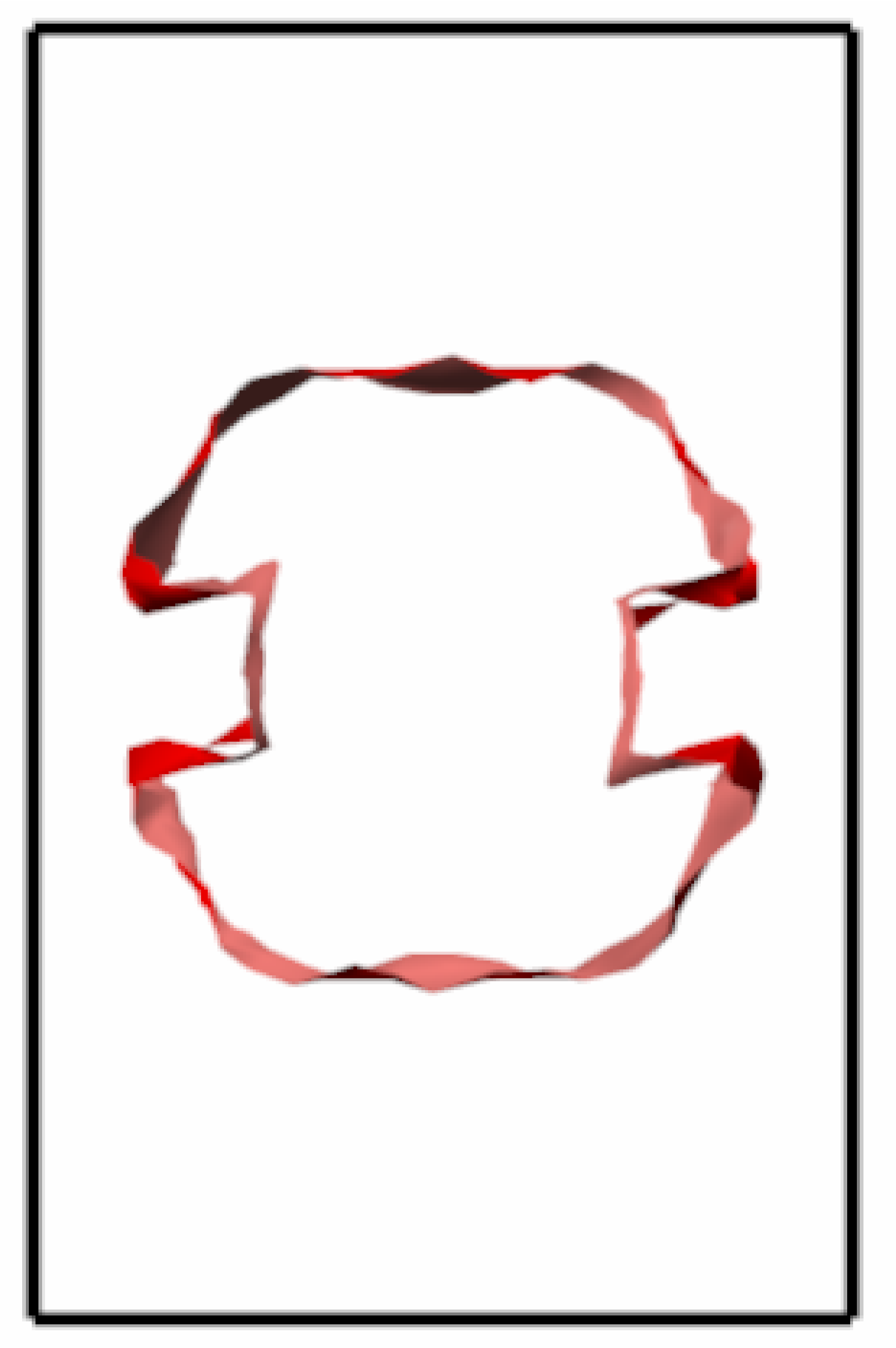}
            \label{fig:band_5}
        }
\hskip 0.1 in
        \subfigure[]{
            \includegraphics[width=\figwii]{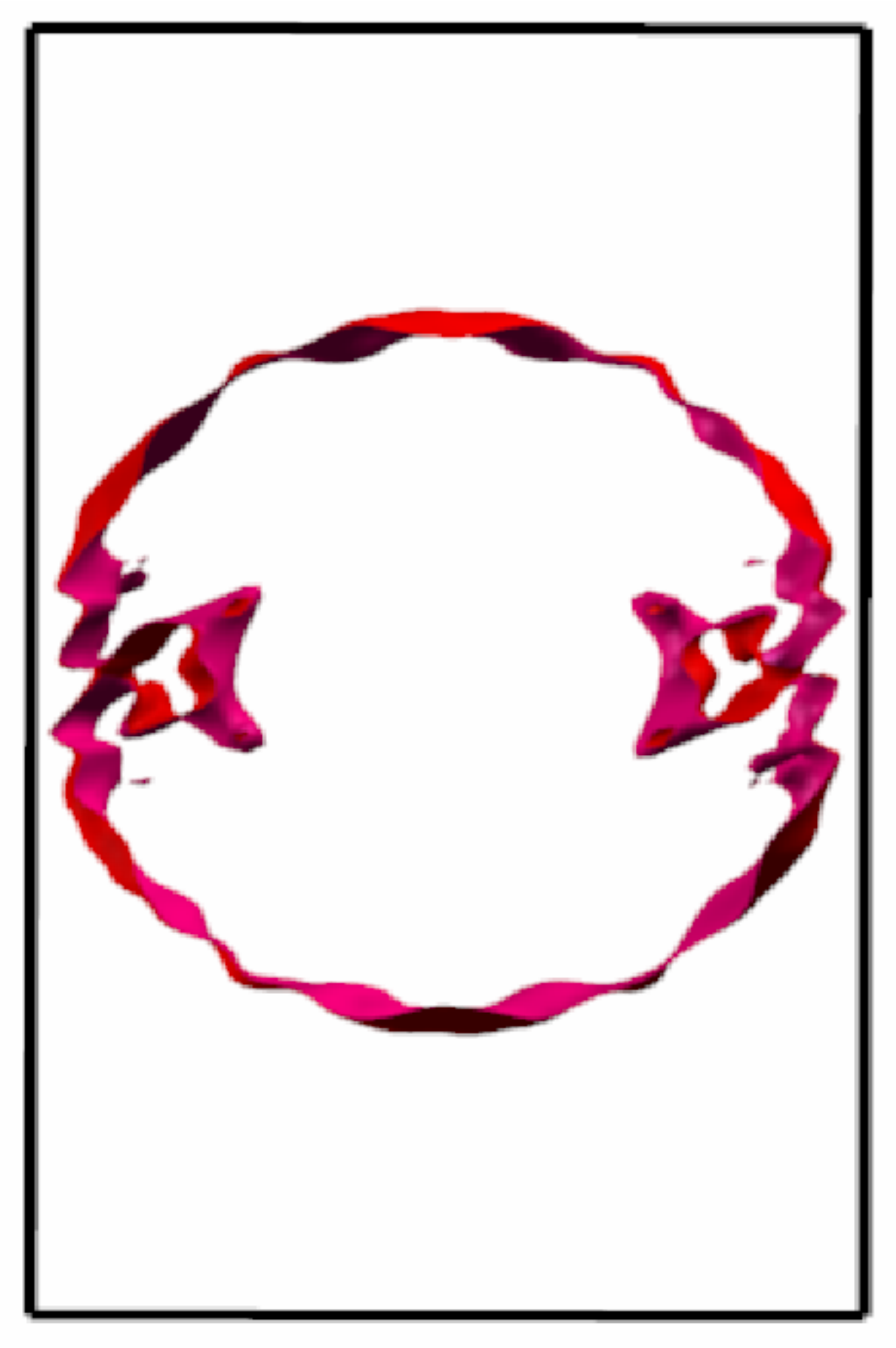}
            \label{fig:band_6}
        }
\hskip 0.1 in
        \subfigure[]{
            \includegraphics[width=\figwii]{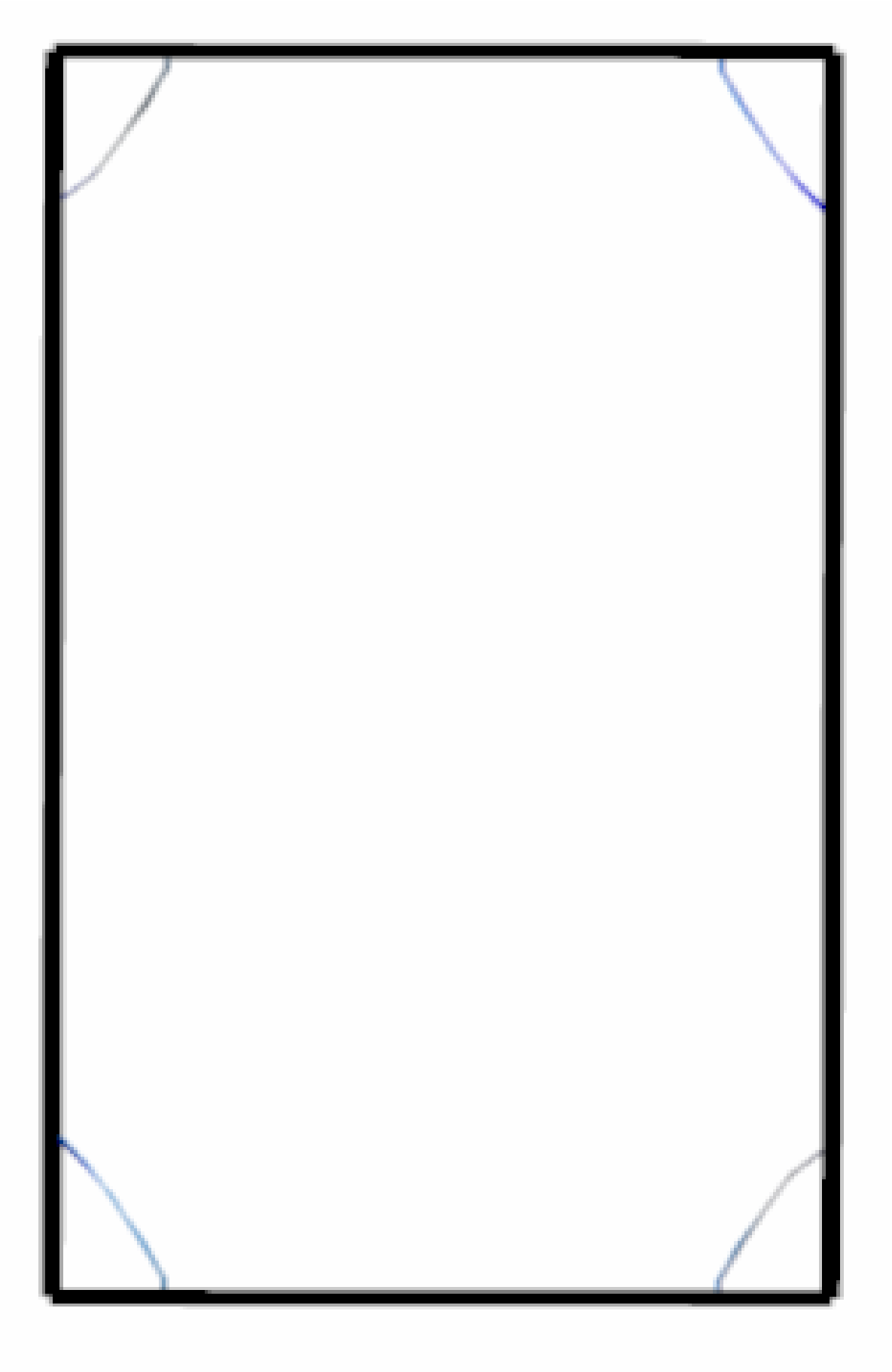}
            \label{fig:band_7}
        }
\hskip 0.1 in
        \subfigure[]{
            \includegraphics[width=\figwii]{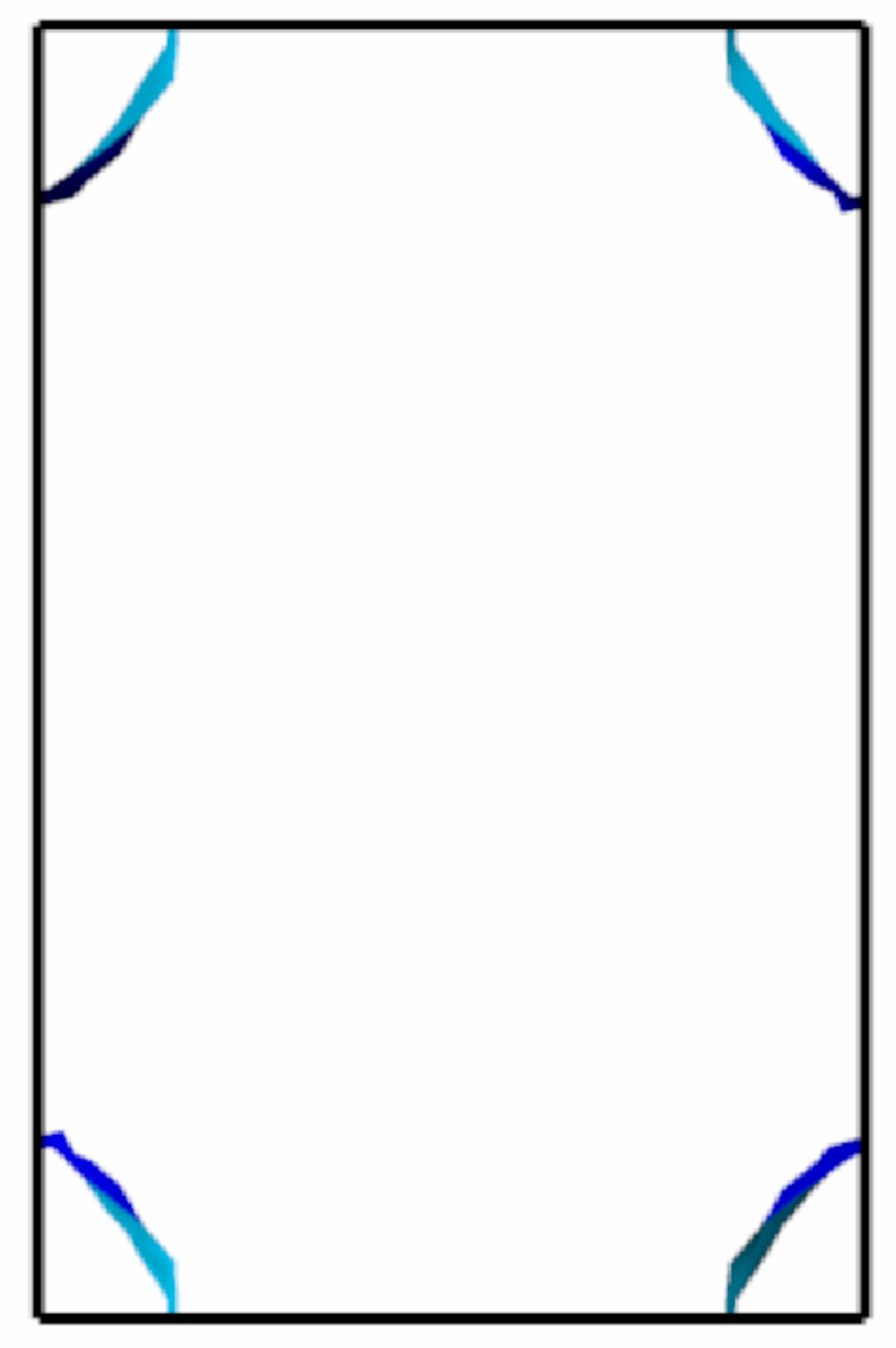}
            \label{fig:band_8}
        }\\

\vskip 0.1 in
        \subfigure[]{
            \includegraphics[width=\figwii]{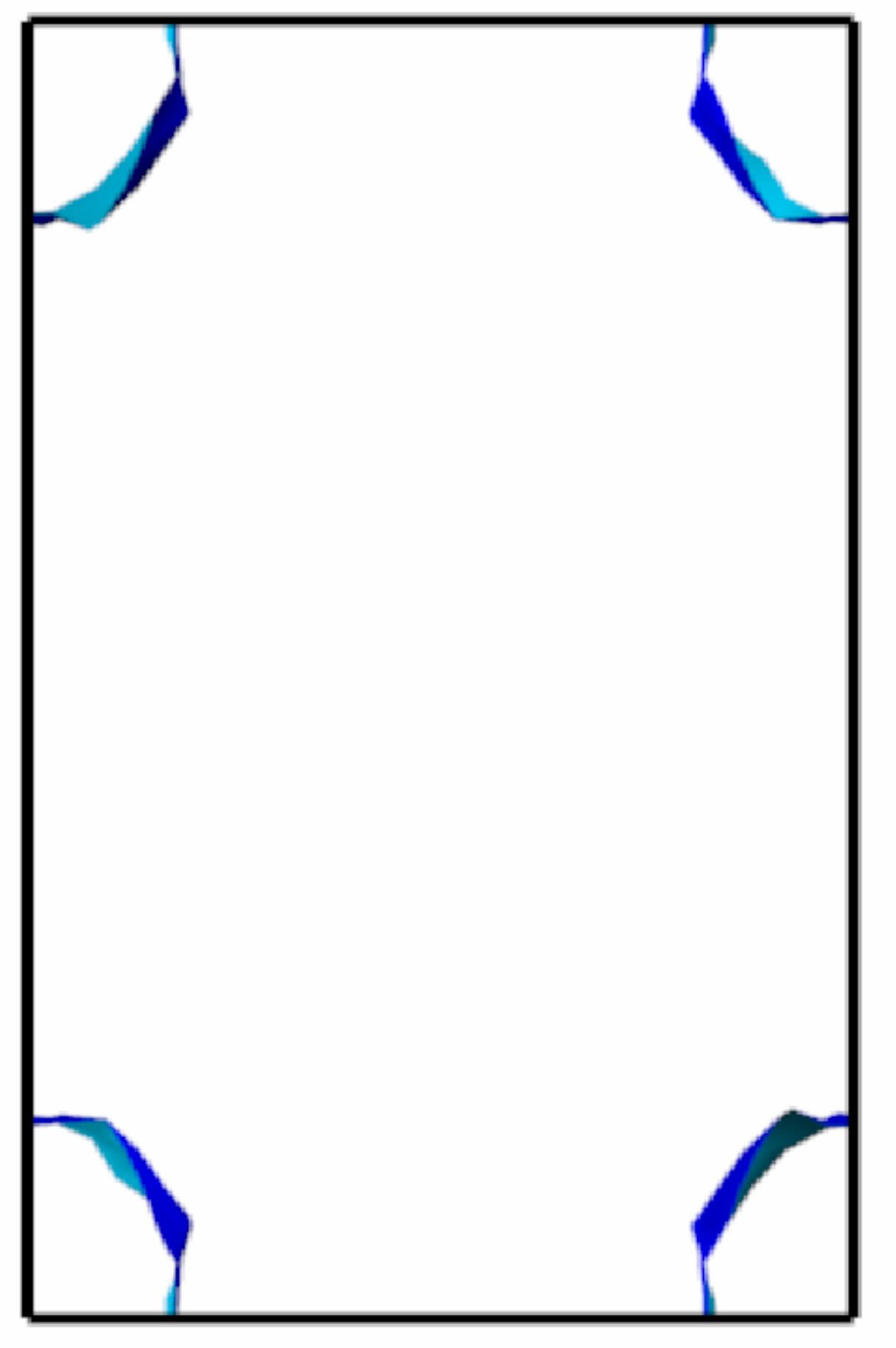}
            \label{fig:band_9}
        }
\hskip 0.1 in
        \subfigure[]{
            \includegraphics[width=\figwii]{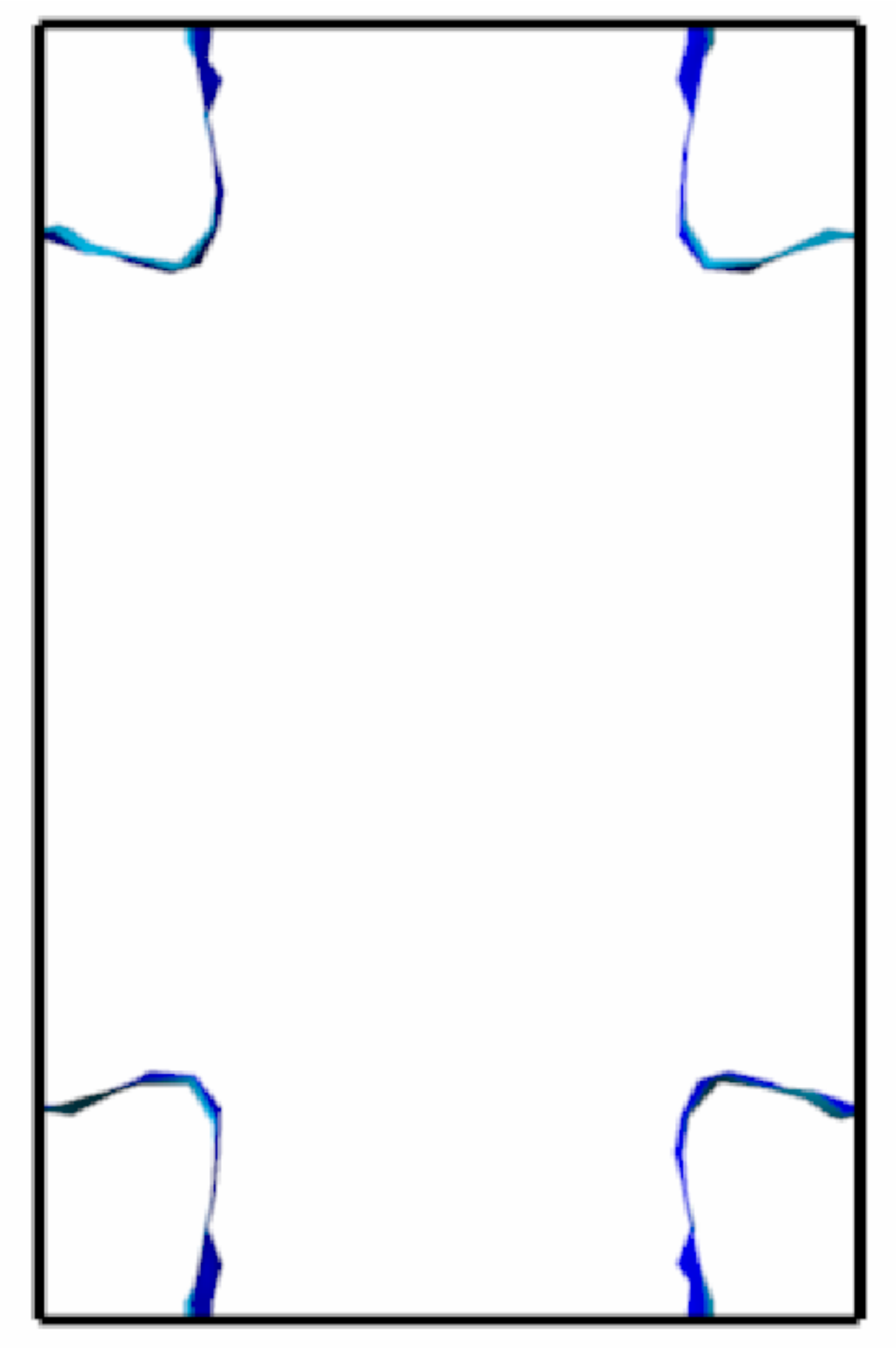}
            \label{fig:band_10}
        }
\hskip 0.1 in
        \subfigure[]{
            \includegraphics[width=\figwii]{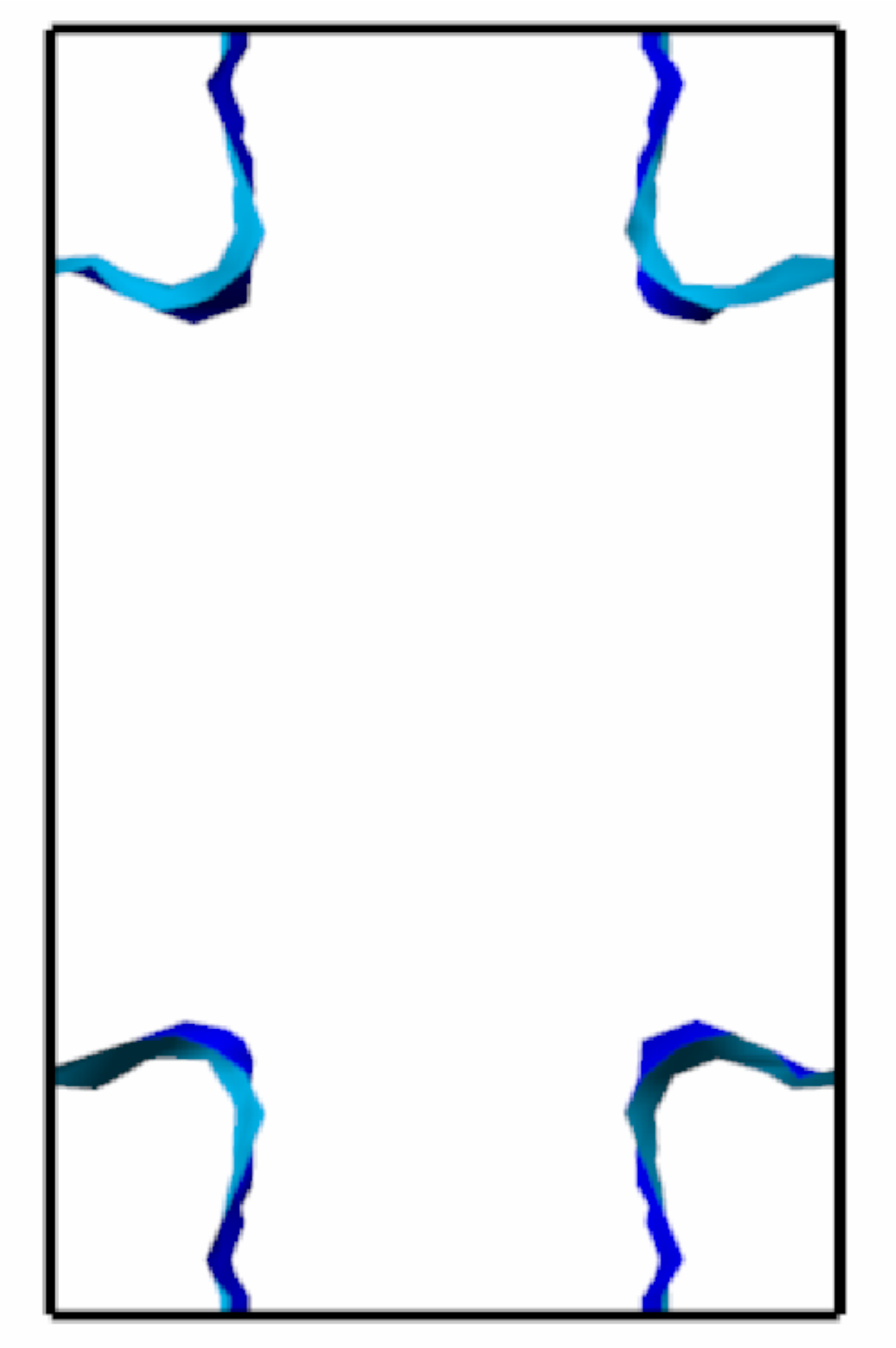}
            \label{fig:band_11}
        }
\hskip 0.1 in
        \subfigure[]{
            \includegraphics[width=\figwii]{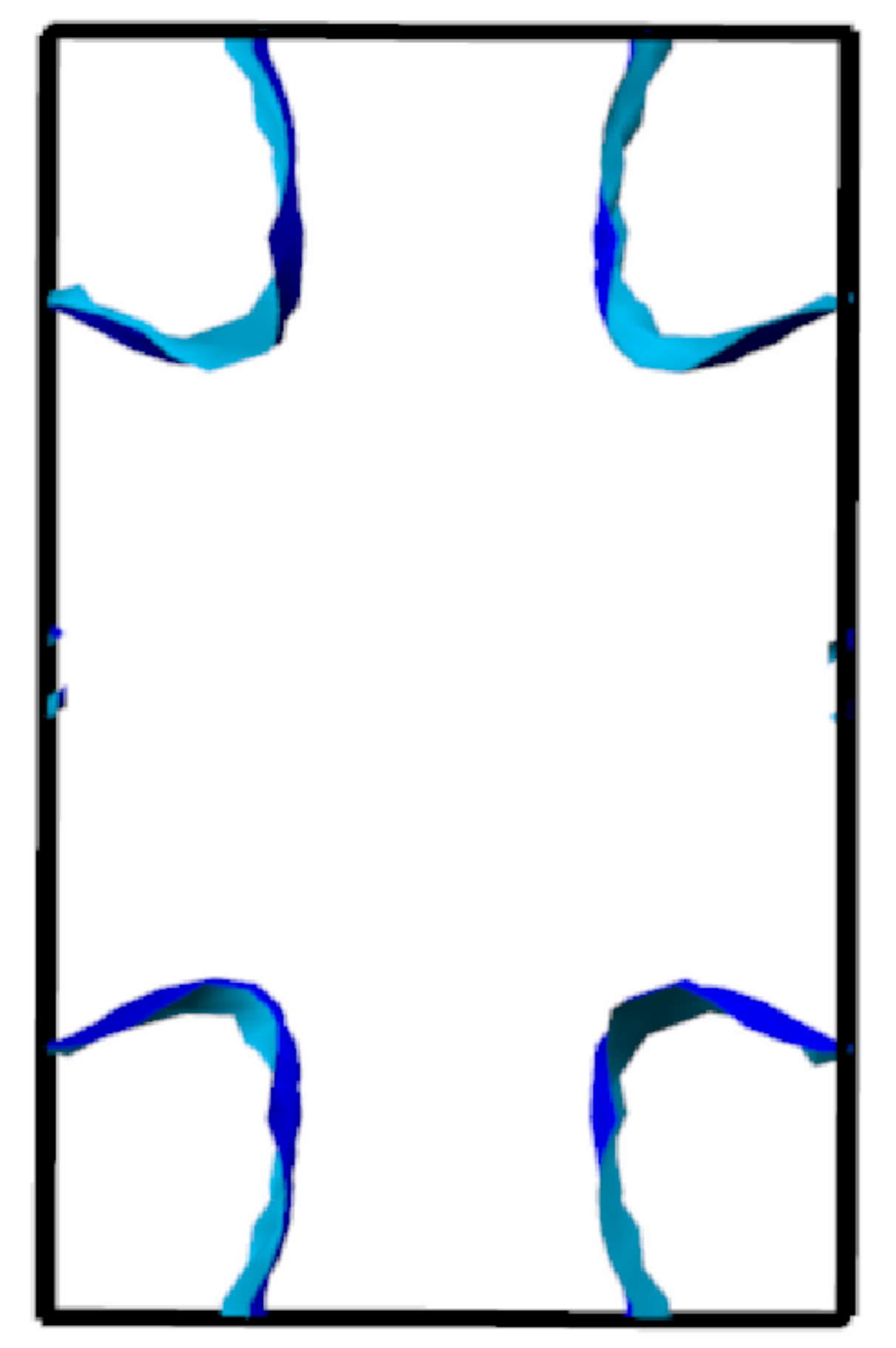}
            \label{fig:band_12}
        }\\

    \end{center}
        \caption{
Fermi surface of CeAl$_4$Ge$_2$Au in the antiferromagnetic phase. (a),(b), (c), (d),(e) and (f) corresponds to hole-pockets
whereas (g),(h),(i), (j),(k) and (l) corresponds to electron sheets.
}
    \label{fig:FS_AF}
\end{figure}

In Fig.~\ref{fig:bands_AF} and Fig.~\ref{fig:FS_AF}, the resulting band structure and Fermi surface of CeAuAl$_4$Ge$_2$ are presented. The bands corresponding to the hole pocket are shown in red whereas the bands corresponding to the electron pockets are shown in blue. Without the U term (Fig ~\ref{fig:CeAlAuGe_nsoc_noU})), there are many flat bands corresponding to Ce-$f$ orbitals close to the Fermi level.
When a non-zero U is applied for two electrons on the Ce-$f$ orbital (Fig.~\ref{fig:bands_AF}(b) for U=6.80 eV, and Fig.~\ref{fig:bands_AF}(b) for J=0.68 eV), we find that the flat bands move away from the Fermi level by an amount of $\approx$ 4 eV, while the location of other bands is robust even for different values of U. Consistent with the experimental quantum oscillation measurements, which probe only the Fermi surface, in this scenario there is no expectation of flat bands with massive electrons near the Fermi level, suggesting that this is an accurate representation of the band structure. In Fig.~\ref{fig:big} we show the projected density of states resulting from this calculation, where the Au, Ge, and Al bands make similar contributions near the Fermi energy.

The calculated Fermi surface (Fig.~\ref{fig:FS_AF}) was  subsequently generated  using  a $k-$point mesh of $17\times 35\times 4$. Because of the doubling of the conventional hexagonal cell only along x-direction, the 2D-BZ becomes rectangular. The calculated frequencies of the extremal orbits on the Fermi surface for magnetic field parallel to the $c$-axis in units of Tesla along with the band masses are gathered in Table ~\ref{table2}, where they are compared to the 5 experimentally observed frequencies. While there were 5 frequencies recorded in the experiment, our calculation shows many frequencies. The smallest and largest calculated frequencies are 33 and 5000 T whereas the experimentally measured ones are 857 and 4500 T respectively. While there is no one-to-one correspondence between the experimental and calculated frequencies, we can group the calculated orbits according to size-similarity as seen in Table~\ref{table2}. When we do that there is semi-quantitative agreement between experiment and calculation. The most significant disagreement is that there are low frequency orbits (below $\sim 300$ T) and a set of orbits with frequency around 2500 T that are absent from the experiment. However, the low frequency orbits, which arise from hole pockets, can be eliminated by pushing the corresponding bands just 4 meV below the Fermi level without affecting the other large frequencies. Since we do not believe that our calculation has this level of accuracy, we find that the only disagreement between our QO experiments and our calculations is the absence of the frequency of the order of 2500 T. We also find that calculated effective masses are in reasonably good agreement with those obtained from experiment.

\begin{table*}[htp]
\begin{tabular}{|p{90pt}|p{50pt}|p{70pt}|p{50pt}|p{50pt}|p{70pt}|} \hline
   Orbits & Calc. freq. & Calc. mass($\mu_e$) & Freq. Label & Expt. freq. & Expt. mass($\mu_e$)    \\ \hline
 ``band-a''  & \textless ~ 20  & & & &    \\ \hline
 ``band-b-pocket 1''  & 33(0.5) & 0.127(0.003) & & & \\
 ``band-f-pocket 1'  & 40 (1) &  0.17(0.01) & & & \\ \hline
 ``band-b-pocket 2''  & 76(7) & 0.16 (0.02) & & & \\
 ``band-b-pocket 3''  & 91(1) & 0.21 (0.01) & & & \\ \hline
 ``band-c-pocket 1'  & 306(2) & 0.29 (0.002)  & & & \\ \hline
 ``band-h-pocket 1'  & 782 (2) & 0.277 (0.001) & & & \\
 ``band-h-pocket 2'  & 833 (2) & 0.262 (0.01) & & & \\
 ``band-g-pocket 1'  & 626 (2) & 0.194 ( 0.002) & ``1''  & 857  & 0.15\\ \hline
 ``band-i-pocket 1'  & 1214 (3) & 0.395 ( 0.002) & ``2''  & 1724 & 0.39\\
 ``band-i-pocket 2'  & 1304 (4) & 0.358 (0.0005) & & & \\
 ``band-c-pocket 2'  & 1265(8) & 0.6(0.1)  &  & &\\ \hline
 ``band-j-pocket 1'  & 1930 (2) & 0.652 ( 0.004) & ``3''  & 1824 & 0.42\\
 ``band-j-pocket 2'  & 1967 (3) & 0.675 (0.002) & & &\\ \hline
 ``band-d-pocket 1'  & 2510(40) & 0.835(0.005) & & & \\
 ``band-k-pocket 1'  & 2445 (5) & 0.81 (0.005) & & & \\
 ``band-k-pocket 2'  & 2584 (4) & 0.75 (0.003) & & &   \\ \hline
 ``band-e-pocket 1'  & 3293 (7) & 1.265 (0.005) & & & \\
 ``band-e-pocket 2'  & 3680 (5) & 1.034 (0.02) & & &  \\
 ``band-l-pocket 1'  & 3608 (5) & 0.79 (0.12) & ``4''  & 3737 & 0.36 \\
 ``band-l-pocket 2'  & 3630 (10) & 0.715 (0.002) & & &  \\
 ``band-l-pocket 3'  & 3680 (7) & 0.92 (0.002) & & & \\ \hline
 ``band-f-pocket 2'  & 5000 (5) & 1.40(0.2)  & ``5'  & 4570 & 0.43  \\ \hline

\end{tabular}
\caption{The frequency of the orbits in Tesla for different pockets in the antiferromagnetic phase. The right panel consists of experimental frequencies and masses which are put next to the nearest matching calculated ones. }
\label{table2}
\end{table*}

\section{\label{sec:level1} DISCUSSION AND CONCLUSION}
CeAuAl$_4$Ge$_2$ provides a useful starting point for investigating geometric magnetic frustration in a cerium based compound without strong Kondo hybridization. This is evidenced by electrical transport and thermodynamic measurements which reveal metallic transport without Kondo coherence, Curie-Weiss behavior (i.e., trivalent cerium), crystal electric field splitting that leads to antiferromagnetic exchange at low temperatures, and magnetic ordering with an antiferromagnetic character at $T_{\rm{M}}$ $=$ 1.4 K. Weak magnetic frustration is further suggested by the presence of magnetic fluctuations well above $T_{\rm{M}}$. Electronic structure calculations reveal that allowing for the experimentally observed antiferromagnetic order and inclusion of the on $f$-site Coulomb repulsion (Hubbard) U causes the $f$-electron bands to move away from the Fermi level, resulting in electronic behavior that is dominated by the $s$-, $p$-, and $d$- bands, which are all characterized by light electron masses. Finally, there is good agreement between the calculated Fermi surface and associated quasiparticle effective masses and results from measured quantum oscillations in ac-magnetic susceptibility measurements. Besides giving a complete picture of the electronic and magnetic behavior of this compound, this result sets the stage to consider systematically exploring the surrounding chemical phase space, with the aid of electronic structure calculations, in order to seek ways to amplify the low temperature antiferromagnetic exchange interaction and thereby induce magnetic frustration in what might be a model family of materials.

\section{\label{sec:level1}ACKNOWLEDGMENTS}
This work was performed at the National High Magnetic Field Laboratory (NHMFL), which is supported by National Science Foundation Cooperative Agreement No. DMR-1157490, the State of Florida and the DOE. Research of RB, KH, YL, and DG supported in part by the Center for Actinide Science and Technology, an Energy Frontier Research Center funded by the U.S. Department of Energy (DOE), Office of Science, Basic Energy Sciences (BES), under Award Number DE-SC0016568. We acknowledge Stan Tozer for use of his 16T PPMS which is also partially supported as part of the above mentioned CAST effort.

\section{Citations}

\newpage
\bibliography{CeAuAl4Ge2}

\end{document}